\documentclass[12pt]{article} 
 
\usepackage{epsf,overcite} 
 
\def \btau{ {{\mbox {\boldmath $\tau$}}}} \def \br{ {  {\mbox 
{\boldmath $r$}} }} \def \bQ{ { {\mbox {\boldmath $Q$}} }}    \def \bu{ { {\mbox {\boldmath $1$}}}} \def \bv{ { {\mbox 
{\boldmath $v$}} }}  \def 
\btau{ {{\mbox {\boldmath $\tau$}}}} \def \bomega{ {{\mbox {\boldmath 
$\omega$}}}} \def \bgam{ {{\mbox {\boldmath $\gamma$}}}} \def \beps{ 
{{\mbox {\boldmath $\epsilon$}}}}  \def \bk{ 
{\mbox {\boldmath $\kappa$}}} \def \avel { {\bigl\langle}} \def \aver { 
{\bigr\rangle}}    \def \bF{ { {\mbox {\boldmath $F$}} }} \def \bFnu{ 
{{\bF}_{\nu}^{(\phi)} }} \def \dphi{ {\partial \phi \over \partial 
{\bQ} } } \def \mom{ {\bigl\langle  \bQ \bQ \bigr\rangle}} \def \bPi{ 
{{\mbox {\boldmath $\Pi$}} }} \def \moms { {\bigl\langle  \bQ^* \bQ^* 
\bigr\rangle}} \def \bGF{ {{\mbox {\boldmath $G$}} }} 
\def \bW{ {{\mbox {\boldmath $W$}} }} 
\def \bE{ { {\mbox {\boldmath $E$}} }}
\def \bY{ { {\mbox {\boldmath $Y$}} }}
 
\makeatletter \renewcommand\@biblabel[1]{$({#1})$} \makeatother
 
\title{\bf {Viscometric Functions for a Dilute Solution of Polymers in 
a Good Solvent}} \author {{\bf J. Ravi Prakash}\\{\it 
Department~of~Chemical~Engineering,}\\ {\it 
Indian~Institute~of~Technology, Madras, India 600 036}\\ \\ {\bf Hans 
Christian \"{O}ttinger}\\{\it Department of Materials, Institute of 
Polymers,}\\{\it ETH Z\"{u}rich, Z\"{u}rich, Switzerland, CH-8092} } 
 
\begin{document} 
\maketitle 
 
\begin{abstract} 
A dilute polymer solution is modeled as a suspension of 
non-interacting Hookean dumbbells and the effect of excluded volume is 
taken into account by incorporating a narrow Gaussian repulsive 
potential between the beads of each dumbbell. The narrow Gaussian 
potential is a means of regularising a delta-function potential---it 
tends to the delta-function potential in the limit of the 
width parameter $\mu$ going to zero. 
Exact predictions of viscometric functions in simple 
shear flow are obtained with the help of a retarded motion expansion 
and by Brownian dynamics simulations. It is shown that for relatively 
small {\it non-zero} values of $\mu$, the presence of excluded volume 
causes a {\it swelling} of the dumbbell at equilibrium, and {\it shear 
thinning} in simple shear flow.  On the other hand, a delta function 
excluded volume potential does not lead to either swelling or to shear 
thinning.  Approximate viscometric functions, obtained by assuming that 
the bead-connector vector is described by a Gaussian non-equilibrium 
distribution function, are found to be accurate above a threshold value 
of $\mu$, for a given value of the strength of excluded volume 
interaction, $z$. A first order perturbation expansion 
reveals that the Gaussian 
approximation is exact to first order in $z$. The predictions of an 
alternative quadratic excluded volume potential suggested earlier 
by Fixman (J. Chem. Phys., 1966, 45, 785; 793) are also compared with 
those of the narrow Gaussian potential.

\end{abstract} 
\thispagestyle{empty}
\newpage
\section {\bf {Introduction }} The fact that two parts of a polymer 
chain cannot occupy the same place at the same time due to their finite 
volume has been recognised in the polymer literature for many years now 
as being an extremely important microscopic phenomenon that governs the 
macroscopic behavior of polymer solutions~\cite{yama, doi, declos}. 
Like hydrodynamic interaction, the {\it excluded volume} effect 
influences the properties of polymer solutions even in the limit of 
extremely long chains because it is responsible for segments remote 
from each other along the polymer chain interacting with each other. 
 
While the effect of excluded volume on static properties of polymer 
solutions has been widely studied, there have been very few attempts at 
examining its influence on properties far from equilibrium. Excluded 
volume effects can be incorporated into bead-spring chain models for 
polymer solutions in a relatively straight\-forward manner by adding 
the excluded volume interaction force between a particular bead and all 
the other beads in the chain (pairwise) to the other potential forces 
that are acting on the bead. An noteworthy aspect of this approach is 
the kind of repulsive potential that is adopted to represent the 
excluded volume interactions.  In static theories of polymer solutions, 
the excluded volume interaction is typically assumed to be a very short 
range $\delta$-function potential. 
 
Fixman~\cite{fix} and more recently Ahn~et al.\ \cite{ahn} have 
attempted to predict the rheological properties of dilute polymer 
solutions by approximately incorporating the effects of both 
hydrodynamic interaction and excluded volume in a self-consistent 
manner into a bead-spring chain model. (Ahn~et al.\ also include 
finitely extensible springs in place of Hookean springs).  In order to 
obtain a solvable model, Fixman~\cite{fix} used a repulsive  
quadratic excluded volume potential in place 
of a $\delta$-function potential. This leads to a tractable  
model since the 
bead-connector vectors are then described by a Gaussian non-equilibrium 
distribution function.  Results obtained with the quadratic 
excluded volume potential have, however, not been compared so far  
with the results of other models for the excluded volume potential. 
 
Andrews~et al.\ \cite{andrews} have recently carried out a 
numerical study of the influence of excluded volume interactions 
on rheological and rheooptical properties of dilute solutions, with the help 
of Brownian dynamics and 
configuration biased Monte Carlo simulations. A bead-spring chain 
model, with ``Fraenkel'' springs between beads and a Morse potential 
to represent 
excluded volume interactions, was used to to model the flexible polymer 
molecule. Attention was largely confined to the prediction of properties 
in elongational flow and transient shear flow.   
 
The predictions of their theories in the limit of long chains have not  
been considered by Fixman~\cite{fix}, Ahn~et al.\ \cite{ahn} 
and Andrews~et al.\ \cite{andrews}.   
On the other hand, the universal character of excluded volume effects  
have been studied using renormalisation group theory methods based on  
kinetic theory models (with a $\delta$-function excluded volume  
potential) by \"Ottinger and coworkers~\cite{ottrg, zylkarg}.  
 
While the work of Andrews~et al.\ \cite{andrews} is based on 
Brownian Dynamics simulations, the accuracy of the other 
{\em approximate} treatments of excluded volume cited  
above has not been tested by comparison with Brownian Dynamics  
simulations (which are an ideal tool for testing approximations for  
nonlinear effects). This is in contrast to the situation that exists 
for kinetic theory models that only incorporate hydrodynamic 
interaction effects, where extensive comparisons between the 
exact results of Brownian Dynamics simulations and various 
approximations have been made~\cite{zylkadb, zylkaga}.  
 
It is the purpose of this paper to examine the influence of the 
excluded volume effect on the rheological properties of a dilute 
polymer solution by using a {\it narrow Gaussian potential} to describe 
the excluded volume interactions. Since the 
narrow Gaussian potential tends to the $\delta$-function potential in 
the limit of a parameter $\mu$ (that describes the width of the 
potential) going to zero, it provides a means of evaluating results 
obtained with a  
singular $\delta$-function potential. Compared to the $\delta$-function 
potential, analytical calculations are not significantly harder with 
the narrow Gaussian potential; quite often, upon setting $\mu=0$ at the 
end of a calculation, the predictions of a $\delta$-function potential 
can be obtained. Furthermore, since Brownian dynamics simulations 
cannot be performed with a $\delta$-function potential, simulations 
carried out with the narrow Gaussian potential for small values of the 
parameter $\mu$ provide a means of asymptotically obtaining the 
predictions of a $\delta$-function potential model. 
 
Any molecular theory that seeks to describe the dynamics of 
polymers in good solvents must simultaneously incorporate both 
the microscopic phenomena of hydrodynamic interaction and excluded 
volume, since hydrodynamic interaction effects have been shown to have 
an unavoidable influence on the dynamic behavior of polymer solutions. 
However, it would be  
difficult to explore the consequences of such a theory for two 
reasons. Firstly, the incorporation of hydrodynamic interaction would 
lead to the complication of multiplicative noise.
Secondly, since Brownian dynamics simulations for long chains
would be extremely computationally intensive, any approximations 
that are developed can only be tested for very short chains. 
For these reasons, and being in the nature of a preliminary 
investigation, we examine excluded volume effects independently from 
hydrodynamic interaction effects, and confine attention here to a 
Hookean dumbbell model for the polymer. This enables 
the careful evaluation of various approximations. It is hoped that,  
in the future, the best approximation can be used for chains with 
both hydrodynamic interaction and excluded volume, with the ultimate 
objective of calculating universal properties predicted in the 
long chain limit.

It must be noted that unlike in the case of the quadratic potential,  
the use of the narrow Gaussian potential does not lead to exact solvability. 
Indeed, as has been observed in earlier treatments of other non-linear 
microscopic phenomenon such as hydrodynamic interaction and internal 
viscosity, it is found that it is not possible to obtain an analytical 
solution valid at all shear rates. As a result, retarded motion 
expansions, Brownian dynamics simulations, and perturbative and 
non-perturbative approximation procedures are used in this work to 
obtain the material functions predicted by the narrow Gaussian potential. 
 
An important consequence of using the narrow Gaussian potential is that 
the nature of Fixman's quadratic excluded volume potential 
can be explored. Predictions of Fixman's quadratic excluded volume 
potential in simple shear flow will be compared with 
the predictions of the narrow Gaussian potential. 
 
The outline of this paper is as follows. In section~2, the basic 
equations required to discuss the dynamics of Hookean dumbbells in good 
solvents are derived. In section~3, the various material functions  
considered in this paper are defined. A retarded 
motion expansion for the stress tensor, for arbitrary excluded volume 
potentials, is derived in section~4, and power series expansions for 
the material functions in simple shear flow are 
obtained.  Section~5 is devoted to examining the consequences of 
describing the excluded volume interaction with a narrow Gaussian 
potential.  In section~5.1, expressions for the material functions 
predicted with this potential at zero shear rate are obtained by using 
the retarded motion expansion. The formulation of a Brownian dynamics 
simulation algorithm is discussed in section~5.2, and a Gaussian 
approximation for the configurational distribution function is 
introduced in section~5.3. In section~5.4, a first order perturbation 
expansion in the strength of the excluded volume interaction is 
derived. Fixman's theory for dumbbells is presented in section~6, 
in terms of the framework adopted for the rest of the 
discussion in this paper. The results of the various exact and 
approximate treatments are compared and discussed in section~7, and the 
principal conclusions of the paper are summarised in section~8.

\section{Basic Equations} The Hookean dumbbell model represents a 
macromolecule by a mechanical model that consists of two identical 
beads connected by a spring.  The solvent in which the beads are 
suspended is assumed to be Newtonian, and attention is restricted to 
flows which have a homogeneous velocity field, {\it ie.} of the form $ 
\bv = \bv_0 + \bk (t) \cdot \br $, where $\bv_0$ is a constant vector, 
$\bk(t)$ is a traceless tensor, and $ \br$ is the position vector with 
respect to a laboratory-fixed frame of reference. The instantaneous 
position of the beads are specified by bead position vectors $ \br_1$ 
and $ \br_2$. 
 
This paper examines the consequence of introducing an excluded volume 
interaction between the beads of the dumbbell, so that the total 
potential experienced by the beads, $ \phi $, is the sum of the spring 
potential $S$, and the excluded volume potential $E$. The force on bead  
$\nu$ due to this potential, $\bFnu$, is then given by  
$ \bFnu = - \, {\partial \phi / \partial \br_{\nu} }$.  
For Hookean springs, the spring potential is given by $S 
= ({1 / 2})\, H Q^2$, where $H$ is the spring constant. Two forms of 
the excluded volume potential are considered in this work, namely, the 
narrow Gaussian potential, and Fixman's quadratic 
potential.  These are discussed in greater detail subsequently. Here,  
we summarise the governing equations that are valid for an 
arbitrary choice of the excluded volume potential $E$. 
 
For homogeneous flows, the configurational distribution function $ 
\psi(\bQ, t )$ depends only on the internal configuration of the 
dumbbell, specified by the bead-connector vector $ \bQ = \br_{2} - 
\br_1 $, and not on the center of mass. The quantity $ \psi( \bQ, t )\, 
d \bQ $ is then the probability that at time $t$ the dumbbell has a 
configuration in the range $\bQ$ to $\bQ + d\bQ$.  Using the framework 
of polymer kinetic theory~\cite{bird2} one can show that the 
distribution function $ \psi( \bQ, t )$, in the presence of excluded 
volume, satisfies the following {\it diffusion equation}, 
\begin{equation} {\partial \psi \over \partial t} = - \, {\partial 
\over \partial \bQ} \cdot \biggl\lbrace (\bk \cdot \bQ) \, \psi - {2 
\over \zeta} \, \psi \, \dphi -  {2 k_B T \over \zeta} \, {\partial 
\psi \over \partial \bQ } \biggr\rbrace \label{diff} \end{equation} 
where, $\zeta$ is the bead friction coefficient ({\it ie.,} $\zeta=6 \pi 
\eta_s a$ for spherical beads with radius $a$, in a solvent with 
viscosity $\eta_s$), $ k_B$ is Boltzmann's constant, and $T$ is the 
absolute temperature. 
 
The stress tensor, $\btau$, in a polymer solution is considered to be 
given by the sum of two contributions, $\btau=\btau^s + \btau^p$, where 
$\btau^s$ is the contribution from the solvent, and $\btau^p$ is the 
polymer contribution. Since the solvent is assumed to be Newtonian, 
$\btau^s= - \, \eta_s \, {\dot \bgam}$, where ${\dot \bgam}$ is the 
rate of strain tensor, ${\dot \bgam} = (\nabla \bv) (t) + (\nabla 
\bv)^\dagger (t)$. The rheological properties of a dilute polymer 
solution may thus be obtained by calculating the polymer contribution 
to the stress tensor, $\btau^p$. For a dumbbell model in the presence 
of excluded volume, it is given by the Kramer's 
expression~\cite{bird2},  
\begin{equation}  
\btau^p=- n \, \avel \, \bQ \, \dphi\,  \aver + nk_BT \, \bu  
\label{kram}  
\end{equation} 
Here, $n$ is the number density of polymers, and angular brackets represent 
averaging with respect to the configurational distribution function 
$\psi (\bQ, t)$. 
 
For both the excluded volume potentials considered here, it will turn 
out that calculation of the second moment $\mom$ is necessary in order 
to evaluate the average in equation~(\ref{kram}). A time evolution 
equation for the second moment can be obtained by multiplying the 
diffusion equation~(\ref{diff}) by $\bQ \bQ$ and integrating over all 
configurations, \begin{equation} {d \over dt} \mom = \bk \cdot \mom + 
\mom \cdot \bk^T + {4 k_B T \over \zeta}\, \bu - {2\over \zeta} \, 
\left[ \, \avel \bQ \dphi \aver + \avel \dphi \bQ \aver \, \right] 
\label{secmom} \end{equation} 
 
It proves convenient for subsequent calculations to introduce the 
following dimensionless variables,  
\begin{equation}  
t^* = {t \over 
\lambda_H}, \quad \bQ^* = \sqrt{{ H \over k_B T}} \, \bQ, \quad \bk^* = 
\lambda_H \bk, \quad \phi^* = S^* + E^*  
\label{nondimvar} 
\end{equation}  
where, $ \lambda_H = (\zeta / 4H)$ is the familiar time 
constant, $ S^* = S / k_B T = (1 / 2) \, {\bQ^*}^2 $ and $E^* = E / k_B 
T $ are the non-dimensional Hookean spring potential and the 
non-dimensional excluded volume potential, respectively. Note that   
$\theta$-solvent values of a typical time scale ($ \lambda_H$) 
and a typical length scale ($\sqrt{k_B T / H}$) are used for the purpose 
of non-dimensionalisation, regardless of the form of the excluded 
volume potential. 
 
The intramolecular force is expected to be in the direction of the 
bead-connector vector. Therefore, it can be written in terms of 
non-dimensional variables as, \begin{equation} {\partial \phi^* \over 
\partial \bQ^*} = H^*(Q^*) \, \bQ^* \label{evforce} \end{equation} 
where $H^*(Q^*)$ is an arbitrary function of the magnitude of $\bQ^*$. 
As a result, the diffusion equation~(\ref{diff}), 
in terms of non-dimensional variables, is given by, 
\begin{equation} {\partial \psi \over \partial t^*} = - \, {\partial 
\over \partial \bQ^*} \cdot \biggl\lbrace \bk^* \cdot \bQ^* - {1 \over 
2} \,  H^*(Q^*) \, \bQ^* \biggr\rbrace \, \psi \, + \, {1 \over 2} \, 
{\partial \over \partial \bQ^*} \cdot {\partial \psi \over \partial 
\bQ^* } \label{nondimdiff} \end{equation} while, the Kramers 
expression~(\ref{kram}) assumes the form, \begin{equation} {\btau^p 
\over nk_BT} = - \avel H^* (Q^*)  \, \bQ^* \bQ^* \aver + \bu 
\label{nondimkram} \end{equation} and the second moment 
equation~(\ref{secmom}) becomes, \begin{equation} {d \over dt^*} \avel 
\bQ^* \bQ^* \aver = \bk^* \cdot \avel \bQ^* \bQ^* \aver + \avel \bQ^* 
\bQ^* \aver \cdot {\bk^*}^T - \avel H^* (Q^*)  \, \bQ^* \bQ^* \aver + 
\bu \label{nondimsecmom} \end{equation} Equation~(\ref{nondimsecmom}) 
is in general not a closed equation for the second moments since it 
depends on the form of the function $H^*(Q^*)$ ({\it ie.} on the choice 
of excluded volume potential). 
 
On examining equations~(\ref{nondimkram}) and~(\ref{nondimsecmom}), it 
is straightforward to see that the second moment 
equation~(\ref{nondimsecmom}) is nothing but the Giesekus expression 
for the stress tensor, \begin{equation} {\btau^p \over nk_BT} = {d 
\over dt^*} \avel \bQ^* \bQ^* \aver - \bk^* \cdot \avel \bQ^* \bQ^* 
\aver - \avel \bQ^* \bQ^* \aver \cdot {\bk^*}^T \label{Giesekus} 
\end{equation} 
 
While the equations derived in this section are valid for arbitrary 
homogeneous flows, in this paper we confine attention to the prediction of 
rheological properties in simple shear flows, defined in the 
section below. 
 
\section{Simple Shear Flows}  
\subsection{Steady simple shear flow}  
Steady simple shear flows are described by a tensor $\bk$ which 
has the following matrix representation in the laboratory-fixed 
coordinate system, \begin{equation} \bk={\dot \gamma} \, \pmatrix{ 0 & 
1 & 0 \cr 0 & 0 & 0 \cr 0 & 0 & 0 \cr } \label{ssf1} \end{equation} 
where ${\dot \gamma}$ is the constant shear rate. 
 
 The three independent material functions used to characterize such 
flows are the viscosity, $\eta_p$, and the first and second normal 
stress difference coefficients, $\Psi_1 \,\,{\rm and}\,\, \Psi_2$, 
respectively.  These functions are defined by the following relations 
\cite{bird1}, \begin{equation} \tau_{xy}^p = - {\dot \gamma}\, \eta_p 
\, ; \quad \quad \tau_{xx}^p- \tau_{yy}^p = - {\dot \gamma^2}\, \Psi_1 
\, ; \quad \quad \tau_{yy}^p- \tau_{zz}^p = - {\dot \gamma^2}\, \Psi_2 
\label{sfvis} \end{equation} 
 
\subsection{Small amplitude oscillatory shear flow} A transient 
experiment that is used often to characterise polymer solutions is 
small amplitude oscillatory shear flow, where the tensor $\bk(t)$ is 
given by, \begin{equation} \bk(t)={\dot \gamma}_0 \, \cos \, \omega t 
\pmatrix{ 0 & 1 & 0 \cr 0 & 0 & 0 \cr 0 & 0 & 0 \cr } \label{usf3} 
\end{equation} Here, ${\dot \gamma_0}$ is the amplitude, and $\omega$ 
is the frequency of oscillations in the plane of flow.  The polymer 
contribution to the shear stress, $\tau_{yx}^p$, depends on time 
through the relation \cite{bird1}, \begin{equation} \tau_{yx}^p=- 
\eta^\prime(\omega)\, {\dot \gamma}_0 \, \cos \, \omega t - 
\eta^{\prime\prime}(\omega)\, {\dot \gamma}_0 \, \sin \, \omega t 
\label{usf4} \end{equation} where $\eta^\prime$ and $ 
\eta^{\prime\prime}$ are the material functions characterising 
oscillatory shear flow. They are represented in a combined form as the 
complex viscosity, $\eta^* =\eta^\prime - i \,\eta^{\prime\prime}$. 
 
In the linear viscoelastic flow regime, the stress tensor is described 
by the linear constitutive relation, 
\begin{equation} 
\btau^p (t) = - \, 
\int_{- \infty}^t d\!s \, G(t-s)\, {\dot \bgam}(s) 
\label{usf5} 
\end{equation} 
where $G(t)$ is the relaxation modulus. As a result, for 
oscillatory shear flows with a small amplitude ${\dot \gamma_0}$, 
expressions for the real and imaginary parts of the complex viscosity 
can be found in terms of the relaxation modulus from the expression, 
\begin{equation} \eta^*= \int_0^\infty G(s)\, e^{-i \omega s} \, d\!s 
\label{usf6} \end{equation} 
 
Note that the zero shear rate viscosity $\eta_{p,0}$ and the zero shear 
rate first normal stress difference $\Psi_{1,0}$, which are linear 
viscoelastic properties, can be obtained from the complex viscosity in 
the limit of vanishing frequency, \begin{equation} \eta_{p,0} = 
\lim_{\omega\to 0} \, \eta^{\prime} (\omega) \, ; \quad \quad 
\Psi_{1,0} = \lim_{\omega\to 0} {2 \, \eta^{\prime\prime} (\omega) 
\over \omega} \label{usf8} \end{equation} 
 
\section{Retarded Motion Expansion} 
 
A retarded motion expansion for the stress tensor, derived previously 
for the FENE dumbbell model~\cite{bird2}, can be adapted to the present 
instance by recognising that the FENE spring potential (as indeed any 
choice of excluded volume potential) is but a particular 
example of a connector force potential between the beads of the 
dumbbell.  In this section, we briefly summarise the development of a 
retarded motion expansion for an arbitrary choice of the excluded volume  
potential. Details of the derivation may be found in  
Bird~et al.\ \cite{bird2}. 
 
We seek a solution of the diffusion equation~(\ref{nondimdiff}) whereby 
the configurational distribution function, $\psi (\bQ, t)$, can be 
written as a product of an equilibrium contribution and a flow 
contribution,  
\begin{equation}  
\psi (\bQ, t) = \psi_{\rm eq} (\bQ) \, 
\phi_{\rm fl} \, (\bQ, t)  
\label{prod}  
\end{equation}  
The governing equation for the flow contribution  
$\phi_{\rm fl} \, (\bQ,t)$, 
\begin{equation}  
{\partial \phi_{\rm fl} \over \partial t^*} = - \, 
\biggl\lbrace \, {\partial \phi_{\rm fl} \over \partial \bQ^*} - 
\phi_{\rm fl} \, {\partial \phi^* \over \partial \bQ^*} \, 
\biggr\rbrace \cdot  \bk^* \cdot \bQ^* + {1 \over 2} \, \biggl\lbrace 
{\partial \over \partial \bQ^*} \cdot {\partial  \phi_{\rm fl} \over 
\partial \bQ^* } - 
 {\partial \phi_{\rm fl} \over \partial \bQ^*} \cdot {\partial  \phi^* 
\over \partial \bQ^* } \, \biggr\rbrace  
\label{flowdiff}  
\end{equation} 
can be obtained by substituting equation~(\ref{prod}) into the 
diffusion equation~(\ref{nondimdiff}), and exploiting the fact that the 
equilibrium distribution function is given by, \begin{equation} 
\psi_{\rm eq} (\bQ) = {\cal N_{\rm eq}} \, e^{- \phi^*} \label{eqdist} 
\end{equation} where ${\cal N_{\rm eq}}$ is the normalisation 
constant. 
 
Regardless of the form of the excluded volume potential, at steady state,   
an exact solution to equation~(\ref{flowdiff}) can be found  
for all homogeneous {\it potential} flows~\cite{bird2}. For more  
general homogeneous flows, however, it is necessary to seek a perturbative 
solution. The flow contribution, $\phi_{\rm fl} \, (\bQ,t)$, is 
assumed to be expandable in a power series in the velocity gradients,  
\begin{equation}  
\phi_{\rm fl} \, (\bQ,t) =  1+ \phi_1 + \phi_2 + \phi_3 + \ldots  
\label{flowdistexp}  
\end{equation}  
where $\phi_k$ is of order $k$ in the velocity gradient. 
 
Partial differential equations governing each of the $\phi_k$ may be 
obtained by substituting equation~(\ref{flowdistexp}) into 
equation~(\ref{flowdiff}) and equating terms of like order.  
Following the procedure suggested in Bird~et al.\ \cite{bird2},  
one can judiciously guess the specific 
forms for the functions $\phi_k$ by noting that each of these functions 
must have certain properties. 
The form of the function $\phi_{1}$ which satisfies these requirements is,  
\begin{equation}  
\phi_{1} =  {1\over 2} \, \bQ^* \cdot {\dot \bgam} \cdot \bQ^*  
\label{phi1} 
\end{equation}  
while the form of $\phi_{2}$ can be guessed to be,  
\begin{equation}  
\phi_{2} = {1 \over 
8} \, (\bQ^* \cdot {\dot \bgam} \cdot \bQ^*)^2 - {1 \over 60} \, \avel 
\, {\bQ^*}^4 \, \aver_{\rm eq} \, {\rm tr} \, ( {\dot \bgam} \cdot 
{\dot \bgam} ) + A^* (Q^*) \, \bQ^* \cdot {\dot \bgam} \cdot \bomega 
\cdot \bQ^*  
\label{phi2}  
\end{equation}  
where, $\avel \quad \aver_{\rm eq}$ denotes an average with  
$ \psi_{\rm eq} (\bQ) $,  
$\bomega$ is the vorticity tensor, defined here as   
$\bomega= \bk^* - {\bk^*}^T $, and the scalar function $A^*(Q^*)$  
obeys the following second order differential equation, 
\begin{equation}  
{d^2 \, A^* \over d \, {Q^*}^2} + \biggl( \, {6 \over Q^*} - 
 H^*(Q^*) \, Q^* \, \biggr) \, {d \, A^*  \over d \, {Q^*}} - 2 \,  A^* 
(Q^*) \, H^*(Q^*) = 1  
\label{Aode}  
\end{equation}  
It is difficult to suggest 
boundary conditions for equation~(\ref{Aode}) other than to say that 
the solution must be such that $ \psi (\bQ, t)$ is bounded.  In the 
case of FENE springs, it is possible to explicitly obtain the 
particular solution of a similar second order differential equation. 
 
It is clear from equation~(\ref{Giesekus}) that at steady state, the 
stress tensor can be found once $\avel \bQ^* \bQ^* \aver$ is known. 
The second moment $\avel \bQ^* \bQ^* \aver$ can be found correct to 
second order in velocity gradients by using the power series 
expansion~(\ref{flowdistexp}) for $\phi_{\rm fl} \, (\bQ,t)$,  
and the specific forms for $\phi_{1}$ and $\phi_{2}$  
in equations~(\ref{phi1}) and (\ref{phi2}), respectively.   
This leads to the following expression for 
the stress tensor, correct to third order in velocity gradients, 
\begin{eqnarray}  
- {\btau^p \over nk_BT} &=&  {\lambda_H \over 3} \, 
\biggl( {H \over k_{\rm B} T}  \biggr) \avel Q^2 \aver_{\rm eq}  
\{ {\dot \bgam} \} +  {\lambda_H^2 \over 30} \, \biggl( {H \over k_{\rm 
B} T}  \biggr)^2 \avel Q^4 \aver_{\rm eq} \,  
\bigl\lbrace 2 {\dot \bgam}^2 - ( {\dot \bgam} \cdot \bomega - 
\bomega \cdot {\dot \bgam} ) \bigr\rbrace \nonumber \\  
&+& {\lambda_H^3 \over 105} \, 
\biggl( {H \over k_{\rm B} T}  \biggr)^3 \, \avel Q^6 \aver_{\rm eq} \, 
\{ \, {3 \over 4} \, {\rm tr} \, ( {\dot \bgam} \cdot {\dot \bgam} ) \, 
{\dot \bgam}  - {1 \over 2} \, ( {\dot \bgam}^2 \cdot 
\bomega - \bomega \cdot {\dot \bgam}^2 )  \bigr\rbrace \nonumber \\ 
 &-&  {\lambda_H^3 \over 180} \, \biggl( {H \over k_{\rm B} T}  \biggr)^3  
\, \avel Q^4 \aver_{\rm eq} \, \avel Q^2 \aver_{\rm eq} \, 
 \{ {\rm tr} \, ( {\dot \bgam} \cdot {\dot \bgam} ) \,  
{\dot \bgam} \}  \nonumber \\ &+& 
{\lambda_H^3 \over 30} \, \biggl( {H \over k_{\rm B} T}  \biggr)^2 \, 
\avel Q^4 \, A^* \aver_{\rm eq} \, \bigl\lbrace \, ( {\dot \bgam}^2 
\cdot \bomega - \bomega \cdot {\dot \bgam}^2 ) + \bomega \cdot ( {\dot 
\bgam} \cdot \bomega - \bomega \cdot {\dot \bgam} ) \nonumber \\  
&-& ({\dot \bgam} \cdot \bomega - \bomega \cdot {\dot \bgam} ) \cdot \bomega 
\, \bigr\rbrace  + \ldots  
\label{retmot}  
\end{eqnarray} 
The Cayley-Hamilton theorem has been used to eliminate the term ${\dot 
\bgam}^3$ in equation~(\ref{retmot}), and an isotropic term that does 
not affect the rheological properties has been dropped. 
 
The stress tensor in simple shear flow, for small values of  
the non-dimensional shear rate $\lambda_H \, {\dot \gamma}$,  
can be found by substituting equation~(\ref{ssf1}) for the rate  
of strain tensor, in equation~(\ref{retmot}).  Using the 
definitions of the viscometric functions in equation~(\ref{sfvis}), the 
following power series expansions are obtained,  
\begin{eqnarray} 
{\eta_{p} \over \lambda_H n k_{\rm B} T } &=&  {1 \over 3} \, \biggl( {H 
\over k_{\rm B} T} \biggr) \, \avel Q^2 \aver_{\rm eq} + \biggl\lbrace 
\, {2 \over 15} \,\biggl( {H \over k_{\rm B} T}  \biggr)^2  \, \avel 
Q^4 \, A^* \aver_{\rm eq} + {1 \over 70} \, \biggl(  
{H \over k_{\rm B} T} \biggr)^3 \, \avel Q^6 
\aver_{\rm eq}  \nonumber \\  
&-& {1 \over 90} \, \biggl( {H \over k_{\rm B} T} 
\biggr)^3 \, \avel Q^4 \aver_{\rm eq} \, \avel Q^2 \aver_{\rm eq} \, 
\biggr\rbrace \, (\lambda_H \, {\dot \gamma})^2 + \ldots   
\label{etap} \\ 
{\Psi_{1} \over \lambda_H^2 n k_{\rm B} 
T }  &=& {2 \over 15} \, \biggl( {H \over k_{\rm B} T}  \biggr)^2 \, 
\avel Q^4 \aver_{\rm eq} + \ldots  
\label{Psi1}  
\end{eqnarray}  
Clearly, equation~(\ref{Psi1}) indicates that one must expand to higher 
orders in velocity gradients before the shear rate dependence of the 
first normal stress difference can be obtained.  Zero shear rate 
properties, however, can be obtained from equations~(\ref{etap}) 
and~(\ref{Psi1}).  
 
\section{The Narrow Gaussian Potential} 
 
In the static theory of polymer solutions it is common to represent 
the dimensionless excluded volume potential, between two  
points on the polymer chain separated by a non-dimensional 
distance $\bQ^*$, with the Dirac delta function, 
\begin{equation} 
 E^*\,(\bQ^*) =   (2 \pi )^{3 / 2} z  \,\, \delta \, (\bQ^*)  
\label{delpot}  
\end{equation} 
where,  $ z =  v \, ( \, {H 
/ 2 \pi k_B T} \,)^{3 / 2} $ is a non-dimensional parameter which 
represents the strength of the excluded volume interaction,  
and in which $v$---which has the dimensions of 
volume---is called the `excluded volume parameter'~\cite{doi}.  
The parameter $z$ is frequently used in theories that  
incorporate excluded volume,   
as it is considered to be the appropriate parameter to be used in   
perturbation expansions.     
As mentioned earlier, excluded volume interactions 
are taken into account in this work by means of a narrow Gaussian 
potential~\cite{ottbk}.  The narrow Gaussian potential has the 
following form in terms of non-dimensional variables,  
\begin{equation}  
E^*\,(\bQ^*)= {z \over 
\mu^3 } \,\, \exp \, \biggl(\, - {1 \over 2} \, {{Q^*}^2 \over \mu^2}\, 
\biggr) 
\label{nondimnagpot}  
\end{equation}  
 
It is clear from equation~(\ref{nondimnagpot}) that the non-dimensional 
parameter $\mu$ controls the extent of the excluded volume interaction, 
and as $\mu \to 0$, the narrow Gaussian potential tends to the 
$\delta$--potential.  The narrow Gaussian potential, 
as mentioned earlier, serves as a means of regularising the 
singular $\delta$--potential and consequently, permits the evaluation of 
results obtained with a $\delta$--potential.  
 
With excluded volume interactions described by the narrow Gaussian 
potential, the function $H^* (Q^*)$, which appears in 
equation~(\ref{evforce}) for the non-dimensional force between the 
beads of the dumbbell, is given by, 
 \begin{equation}  
H^* (Q^*) \equiv   H_G^*(Q^*) 
 = 1 - \left( {z \over \mu^5} \right) \, \exp \, 
\left[ \, - \, {{Q^*}^2 \over 2 \mu^2}\, \right] 
\label{nondimevforce}  
\end{equation} 
The complex form of this function implies that the diffusion 
equation~(\ref{nondimdiff}) cannot be solved exactly analytically to 
obtain the non-equilibrium configurational distribution function 
$\psi(\bQ, t )$. Furthermore, the time evolution equation for the 
second moments~(\ref{nondimsecmom}) is not a closed equation for the 
second moments since it involves the higher order moment $\avel H_G^* 
(Q^*)  \, \bQ^* \bQ^* \aver$ on the right hand side.  As a result, 
perturbative methods, non-perturbative approximation procedures or 
numerical schemes must be used in order to obtain the material 
functions predicted by the narrow Gaussian potential. 
 
\subsection{Zero shear rate properties} 
 
The viscosity and the first normal stress 
difference predicted by the narrow Gaussian potential at low shear rates 
can be obtained from the equations~(\ref{etap}) and 
(\ref{Psi1}), respectively, once the equilibrium averages that occur in 
these expressions are evaluated. For the narrow Gaussian potential, the 
equilibrium distribution function is given by equation~(\ref{eqdist}), 
with the excluded volume contribution to the non-dimensional potential 
$\phi^*$ given by equation~(\ref{nondimnagpot}). We denote the various 
non-dimensional moments of the equilibrium distribution for 
a narrow Gaussian potential by, 
\begin{equation}  
q_m  \equiv  \, \biggl( {H \over k_{\rm B} T} 
\biggr)^m \, \avel Q^{2m} \aver_{\rm eq} \, ; \quad m=1,2,3, \ldots 
\end{equation}  
 
In order to obtain the viscosity at non-zero 
shear rates, it is necessary to find the function $A^*$ that 
satisfies the second order differential equation~(\ref{Aode}) [with 
$H^* (Q^*) = H^*_G(Q^*)$].  Unlike in the case of an FENE dumbbell, it 
has not been possible to obtain the particular solution to 
equation~(\ref{Aode}).  As a result, attention is confined here 
to obtaining the 
zero shear rate predictions of the narrow Gaussian potential,  
\begin{eqnarray} 
{\eta_{p,0} \over \lambda_H n k_{\rm B} T } &=& {1 \over 3} \, q_1 
\label{etap02} \\  
{\Psi_{1,0} \over \lambda_H^2 n k_{\rm B} T }  
&=& {2 \over 15} \, q_2  
\label{Psi102}  
\end{eqnarray} 
for which only the moments $q_1$ and $q_2$ are required. 
Alternative methods will be used in sections~5.2 to 5.4 to 
obtain the shear rate dependence of the viscometric functions. 
 
The non-dimensional equilibrium moments $q_1$ and $q_2$ can be obtained 
exactly, as will be shown in section~5.1.1 below.  
The need to calculate equilibrium moments is also frequently 
encountered in static theories of polymer solutions.  
In these theories, as mentioned earlier, it is common to represent  
excluded volume interactions with a delta-function excluded volume  
potential, and furthermore, to obtain universal predictions 
by considering the limit of long 
chains. The Hamiltonian then typically has two singular 
objects---making it impossible to evaluate equilibrium moments exactly. 
The most succesful approach so far towards approximately evaluating 
these moments has been to develop a perturbation expansion in 
the parameter $z$, and to use renormalisation group methods to 
refine the results of the perturbation calculation~\cite{doi,declos}. 
An alternative and simpler non-perturbative  
route is the uniform expansion model~\cite{doi}.  
The use of the narrow Gaussian potential---albeit in the simple  
context of Hookean dumbbells---provides an opportunity to compare 
the results of these approximate models with 
the exact solution. The rigorous solution is discussed in 
section~5.1.1 below, while the perturbation expansion and 
the uniform expansion models are discussed in 
sections~5.1.2 and~5.1.3, respectively.  
 
\subsubsection{Exact solution} 
  
It is straight forward to show that   
the moments $q_m$ are given by the ratio of two integrals,   
$ q_m =  (I_m / I_0)$, where,   
\begin{equation}   
 I_j \equiv \int_0^\infty \,  
{Q^*}^{2j+2} \, \exp \,  \bigl\lbrace \, - \, (1/2) \, {Q^*}^2  -  
E^* \, \bigr\rbrace \, d Q^* \, ;  \quad j=0,1,2,\ldots   
\label{sj}   
\end{equation}  
  
In the limit of $\mu \to 0$ and $\mu \to \infty$, these integrals   
can be evaluated analytically.   
Consider the quantities,  
$$ p_j (Q^*)\equiv  {Q^*}^{2j+2} \, \exp \, \lbrace   
\, - \, {1 \over 2 } \, {Q^*}^2  - E^* \, \rbrace$$  
which are the integrands for the integrals $I_j$, and   
$$R_j (Q^*) \equiv  {Q^*}^{2j+2} \, \exp \, \lbrace   
\, - \, {1 \over 2 } \, {Q^*}^2  \, \rbrace$$   
Now, $p_j(0) = R_j(0) = 0$, for all values of $\mu$.   
At any non-zero value of $Q^*$, it is clear from   
equation~(\ref{nondimnagpot}) that,   
$$p_j (Q^*) \to R_j (Q^*) \quad {\rm as} \quad \mu \to 0   
\quad {\rm or} \quad \infty $$   
  
In other words, for all values of $Q^*$, the quantities $p_j (Q^*)$ tend   
{\it pointwise} to $R_j (Q^*) $ as $\mu$ tends to zero or to infinity.   
Furthermore, for all values of $\mu$, 
it can be shown that $p_j (Q^*)$ are bounded functions 
of $Q^*$ on $[0,\infty[$.    
It then follows from a theorem of the calculus~\cite{buck} that,  
$$ q_m \to { \int_0^\infty \, {Q^*}^{2m+2} \, \exp \,  
\bigl\lbrace \, - \, {1 \over 2 } \, {Q^*}^2 \, \bigr\rbrace \, d Q^*  
\over \int_0^\infty \, {Q^*}^{2} \, \exp \,  \bigl\lbrace \, -   
\, {1 \over 2 } \, {Q^*}^2 \, \bigr\rbrace \, d Q^* }   
\quad {\rm as } \quad \mu \to 0 \quad {\rm or} \quad  \infty $$  
  
As a result, the asymptotic values of $q_m$ for $\mu \to  
0$ and $\mu \to \infty$ are found to be {\it independent} of $z$,   
and are equal to the $\theta$-solvent values,   
$$ q_1=3 \, ; \quad q_2=15 \, ; \quad q_3=105  \, ; \, \ldots$$ 
This implies---from equations~(\ref{etap02}) and~(\ref{Psi102})---that   
the use of a delta-function potential to represent excluded volume 
interactions leads to the prediction of zero 
shear rate properties in good solvents which are identical to those in  
$\theta$-solvents.  
 
Away from these limiting values of $\mu$, {\em i.e}.\ at non-zero   
finite values of $\mu$, the integrals $I_j$   
can be found by numerical quadrature. Here they have been evaluated   
using a routine given in Numerical recipes~\cite{numrec} for the   
integration of an exponentially decaying integrand. Discussion of  
zero shear rate property predictions in this case   
is taken up in section~7.  
  
\subsubsection{Perturbation expansion } 
 
Static theories for polymer solutions indicate that accounting 
for excluded volume interactions with a delta function potential 
leads to the prediction of  
a {\em swelling} ({\em i.e}.\ an increase in the mean square end-to-end 
distance) of the polymer chain, which is in close agreement  
with experimental observations~\cite{doi,declos}.  
In the case of Hookean dumbbell, however, we have seen above that  
the use of a delta function potential to account for the presence  
of excluded volume (which corresponds to the limit $\mu \to 0$),  
does not lead to any change in the prediction of equilibrium moments  
when compared to the $\theta$-solvent case.  
It is worthwhile therefore to examine the nature of the perturbation  
expansion in $z$, and to compare it with the results of the exact calculation.   
  
Upon expanding $e^{-E^*}$ in a power series, the integral $I_j$ has the form,  
 \begin{equation}   
 I_j \equiv \int_0^\infty \,  d Q^* \,   
\sum_{n=0}^{\infty} \, u_n (Q^*) \, ;  \quad j=0,1,2,\ldots   
\label{sjexp}   
\end{equation}  
where,  
\begin{equation}   
 u_n (Q^*) = {(-1)^n \over n! } \, {Q^*}^{2j+2} \,  
e^{- {1 \over 2} \, {Q^*}^2} \, \left( \, E^*(Q^*)\, \right)^n  
\label{un}   
\end{equation}   
In order to carry out a term by term integration of the functional  
series $\sum_{n=0}^{\infty} u_n (Q^*)$ in equation~(\ref{sjexp}), 
it is necessary for the series  to be {\it uniformly 
convergent} on $[0,\infty[$. For all values of $z$, and $\mu \neq 0$, 
uniform convergence can be established with the help of the Weierstrass 
$M$ test~\cite{buck}.  
Therefore, a term by term integration in  
equation~(\ref{sjexp}) can be caried out, {\em except} when $\mu =0$. 
Assuming that $\mu \neq 0$, and performing the 
integration in equation~(\ref{sjexp}), one obtains,  
\begin{equation}   
 I_j  = 2^{j+{1 \over 2}} \, \Gamma \biggl(j+{3 \over 2} \biggr)  
\, \sum_{n=0}^{\infty} \,  
{ (-1)^n \over n! } \biggl({z \over \mu^3} \biggr)^n \, 
{ \mu^{2j+3} \over  (n+\mu^2)^{j+{3 \over 2}}}  
;  \quad j=0,1,2,\ldots   
\label{ijper}   
\end{equation}  
 
Consider the moment $q_1$, which is  undoubtedly the most interesting  
physical moment. Using the perturbation expansion for $I_j$, 
$q_1$ is given by the ratio, $q_1 = 3 \, (S_1 / S_0)$,  
where, $S_0$ and $S_1$ are defined by, 
\begin{eqnarray} 
S_0 &=& \sum_{n=0}^{\infty} \,  
{ (-1)^n \over n! } \biggl({z \over \mu^3} \biggr)^n \, 
{ \mu^{3} \over  (n+\mu^2)^{3 \over 2}}  \nonumber \\ 
S_1 &=& \sum_{n=0}^{\infty} \,  
{ (-1)^n \over n! } \biggl({z \over \mu^3} \biggr)^n \, 
{ \mu^{5} \over  (n+\mu^2)^{5 \over 2}}  
\label{s0s1} 
\end{eqnarray} 
If only the first order perturbation term is retained, one obtains,
\begin{equation}   
q_1 = 3 \, \left( 1+{ z \over (1 + \mu^2)^{5/2} } \right) 
\label{equimom}   
\end{equation}  
Curiously, the limit $\mu \to 0$ can be carried out in this case. It 
leads to a result which is in line with static theories, which indicate a 
finite non-vanishing effect due to the presence of  
$\delta$-potential excluded volume interactions. However, such a  
limit cannot be carried out if higher order terms are 
retained, since both the sums $S_0$ and $S_1$ diverge as 
$\mu \to 0$. In static theories, higher order perturbation 
expansions are obtained by {\em dropping} divergent terms, as they 
are postulated not to matter. These divergent terms arise from products of 
the $\delta$-potential for the {\em same} pair of interacting beads. In 
dumbbells, these are the only kind of beads present. As a result, 
there is no meaningful or mathematically consistent way of going 
beyond first order perturbation theory for dumbells in the limit 
$\mu \to 0$.  
 
Note that both the series, $S_0$ and $S_1$, 
are alternating series. Furthermore, for given values of $z$ and $\mu$, the  
terms decrease monotonically for large enough values of $n$. It follows then,  
from the Leibnitz criterion for alternating series~\cite{arfken}, that  
both $S_0$ and $S_1$ converge for all values of $z$ and $\mu \neq 0$.  
This suggests that, even though it is not possible to switch the 
integral and summation in equation~(\ref{sjexp}) for $\mu = 0$, 
the value of $q_1$ at $\mu = 0$ can be found by setting it equal 
to the limit of $q_1(\mu)$ as $\mu \to 0$. We shall see 
in section~7 that such a limiting process is infeasible since it  
becomes numerically impossible to evaluate the sums~(\ref{s0s1})  
for small enough values of $\mu$.  
 
\subsubsection{Uniform expansion model} 
The uniform expansion model seeks to approximate the average $\avel X 
\aver_{\rm eq}$ of any quantity $X(\bQ)$, with $\avel X \aver_{\rm 
eq}^\prime$, where $\avel \quad \aver_{\rm eq}^\prime $ denotes an 
average with the Gaussian equilibrium distribution function, 
\begin{equation} \psi_{\rm eq}^\prime (\bQ) = {\cal N_{\rm eq}^\prime} 
\, \exp \,  \bigl\lbrace \, - \, {3 \over 2 {b^\prime}^2 } \, Q^2 \, 
\bigr\rbrace \label{uemgau} \end{equation} with ${\cal N_{\rm 
eq}^\prime} = [ \, 3 / 2 \pi {b^\prime}^2 \, ]^{3 / 2}$. The aim is to 
find the parameter $b^\prime$ that leads to the best possible 
approximation. As may be expected, this depends on the quantity 
$X(\bQ)$ that is averaged. The motivation behind the uniform expansion 
model, and details regarding the calculation of $b^\prime$ are given in 
appendix A. Since the equilibrium distribution function in the absence 
of excluded volume is Gaussian, this assumption expects the equilibrium 
distribution to remain Gaussian upon the incorporation of excluded 
volume, albeit with an increased end-to-end vector. 
 
If we define $u_m$, such that $u_m \equiv  \, ( {H / k_{\rm B} T} )^m 
\, \avel Q^{2  m} \aver_{\rm eq}^\prime$, then clearly $u_m$ is the 
uniform expansion model approximation for $q_m $.  The uniform 
expansion model predictions of the zero shear 
rate properties are given by  
(\ref{etap02}) and~(\ref{Psi102}), with $q_m $ replaced by $u_m$. 
Results of material properties obtained by numerical quadrature, and 
by the uniform expansion model, are discussed in section~7. 
 
\subsection{Brownian dynamics simulations} 
 
Development of the retarded motion expansion has proved useful in 
obtaining exact expressions for the zero shear rate properties.  At 
non-zero shear rates, exact predictions of the viscometric functions 
can be obtained by solving the Ito stochastic differential 
equation,  
\begin{equation} 
d \bQ^* = \left[ \, \bk^* - {1 \over 2} \, H^*(Q^*) \, \bu \, \right]  
\cdot \bQ^* \, dt + d \bW  
\label{ito} 
\end{equation} 
which corresponds to the non-dimensional diffusion  
equation~(\ref{nondimdiff}), and in which $\bW$ is a three-dimensional  
Wiener process.  
 
For the narrow Gaussian potential, since $H^* (Q^*) = H^*_G(Q^*)$,  
equation~(\ref{ito}) is non-linear. As a result, it cannot be  
solved analytically.  
Two different Brownian dynamics simulation algorithms have been  
adopted here for the numerical solution of equation~(\ref{ito}).  
Both schemes use a second order
predictor-corrector algorithm with time-step 
extrapolation~\cite{ottbk}. The first scheme obtains
steady-state expectations by the simulation of a single long
trajectory, and is based on the assumption of ergodicity~\cite{ottbk}.
It has been used to obtain results at equilibrium, and for large values
of the shear rate.  A second algorithm---which employs
a variance reduction procedure---has been used at low values of the
shear rate, since the variance for the viscometric functions is found
to be relatively large at these shear rates. Reduction in the variance
is obtained by following a scheme suggested by Wagner and
\"Ottinger~\cite{wagott}. The scheme---which constructs an ensemble of
trajectories over several relaxation times, from start-up of 
flow to steady-state---essentially consists
of subtracting the rheological properties obtained from a parallel
equilibrium simulation. While  
this doesn't change the average values of the properties,  
it significantly reduces the fluctuations, since the fluctuations  
are virtually the same at zero and small shear rates. The results of these 
simulation algorithms are discussed in section~7.  
 
\subsection{The Gaussian approximation} 
 
The main obstacle (in the configuration space of the dumbbell) to 
obtaining the rheological properties predicted by a narrow Gaussian 
potential is that the second moment equation is not a closed equation. 
A closure problem has also been encountered earlier in treatments of 
the phenomenon of hydrodynamic interaction and internal viscosity, 
where it has been shown that an accurate approximation can be obtained 
by assuming that the non-equilibrium configurational distribution 
function is a Gaussian distribution~\cite{zylkadb, zylkaga, 
prakbk, schiebiv, wedgeiv}. In this section, 
a similar systematic 
approximation procedure for the treatment of excluded volume 
interactions described by a narrow Gaussian potential is introduced. 
 
The assumption that $\psi(\bQ,t)$ is a Gaussian distribution, 
\begin{equation} \psi(\bQ,t) = {1 \over (2 \pi )^{3 / 2}} \, {1 \over 
{\sqrt{ {\rm det} \mom}}} \,\, \exp \, \biggl\lbrace \, - {1 \over 2} 
\, \bQ \cdot \mom^{-1} \cdot \bQ \, \biggr\rbrace \label{gausdist} 
\end{equation} makes the second moment equation~(\ref{nondimsecmom}) a 
closed equation, since the higher order moment $\avel H_G^* (Q^*)  \, 
\bQ^* \bQ^* \aver$ can be expressed in terms of the second moment. On 
performing this reduction, it can be shown that the Gaussian 
approximation leads to the following closed second moment equation, 
\begin{equation}  
{d \,  \over dt^* } \moms  =  \bk^* \cdot \moms + \moms  
\cdot {\bk^*}^T - \moms +  {z \over {{\sqrt {{\rm det} \,  
\lbrack \, {\moms} + \, \mu^2 \, {\bu } \rbrack }}}} \, \bPi + \bu  
\label{gasecmom}  
\end{equation}  
where, $$ \bPi = \bigl\lbrack 
\moms + \mu^2 \, {\bu} \, \bigr\rbrack^{\, -1} \cdot  \moms$$ 
 
It is also straight forward to show that on introducing the Gaussian 
approximation, the Giesekus expression for the stress tensor has the 
form, 
\begin{equation} 
{\btau^p \over n k_B T } =  - \moms + {z \over 
{\sqrt {{\rm det} \, \lbrack \, {\moms} + \, \mu^2 \, {\bu } \rbrack 
}}} \, \bPi + \, \bu 
\label{gakram} 
\end{equation} 
The steady state viscometric functions [defined by 
equations~(\ref{sfvis})] can therefore be found once 
equation~(\ref{gasecmom}) is solved for $\moms$. 
 
It is worth examining the nature of the polymer contribution to the 
stress tensor. In the limit $\mu \to 0$, $\bPi$ reduces to $\bu$, and 
as a result, the presence of excluded volume only makes an indirect 
contribution through its influence on the second moment $\moms$.  This 
follows from the fact that an isotropic contribution to the stress 
tensor makes no difference to the rheological properties of the polymer 
solution. On the other hand, for non-zero values of $\mu$, the 
rheological properties are also affected directly by excluded volume. 
 
Under the Gausssian approximation, linear viscoelastic properties  
can be obtained by deriving a first order codeformational memory  
integral expansion for the stress tensor.  
The tensor $\moms$ is expanded, in terms of deviations from its 
isotropic equilibrium solution, upto first order in velocity gradient, 
\begin{equation}  
\moms= \alpha^2 \, ( \bu + \beps + \ldots \, ) 
\label{linvisexp}  
\end{equation} 
where, the parameter $\alpha$ 
(commonly called the swelling ratio) is defined by,  
\begin{equation} 
\alpha^2= {\avel \, Q^2 \, \aver_{\rm eq} \over \avel \, Q^2 \, 
\aver_{0,\rm eq}}  
\label{alpha}  
\end{equation}  
Here, $\avel \, Q^2 \, 
\aver_{0,\rm eq} = ({3 k_B T / H})$, is the mean square end-to-end 
distance in the absence of excluded volume.  Clearly, $\alpha$ 
represents the {\it equilibrium} swelling of the polymer chain caused 
by the presence of excluded volume.  $\alpha$ is not an 
independent parameter since at equilibrium the second moment 
equation~(\ref{gasecmom}) reduces to the following consistency 
relationship between $z$, $\mu$ and $\alpha$, 
\begin{equation} 
z =  (1 - \alpha^{-2} \, ) \, \lbrack \, 
\alpha^2 + \mu^2 \, \rbrack^{5 \over 2} 
\label{equirel} 
\end{equation} 
The well known scaling relation for the end-to-end 
distance with the number of monomers $N$, namely, 
$\sqrt{ \avel \, Q^2 \, \aver_{\rm eq}} \sim N^{3/5}$, 
may be obtained from equations~(\ref{alpha}) and 
(\ref{equirel}) in the limit of large $N$ by noting that---since excluded 
volume is a pairwise interaction---$z$ must scale as $\sqrt{N}$ for 
dumbbells~\cite{larson}. 

Substituting the expansion~(\ref{linvisexp}) into the evolution 
equation~(\ref{gasecmom}) leads to, \begin{equation} {d \over dt^*} \, 
\beps= \bk^* + {\bk^*}^T - {1 \over \tau^*} \, \beps \label{lvdif} 
\end{equation} where, \begin{equation} \tau^* =  \biggl[ 1 - {z \, 
\mu^2 
 \over ( \, \alpha^2 + \mu^2 \, )^{7/2}} \biggr]^{\, -1} \label{lvtau} 
\end{equation} Furthermore, the stress tensor~(\ref{gakram}) upto first 
order in the velocity gradient (without the rheologically unimportant 
isotropic contribution) is given by, \begin{equation} {\btau^p \over n 
k_B T } =  - {\cal H } \, \beps \label{lvkram} \end{equation} where, 
\begin{equation} {\cal H} = \alpha^2 \, ({\tau^*})^{\, -1} \label{lvH} 
\end{equation} 
 
Upon integrating equation~(\ref{lvdif}), which is a first order 
ordinary differential equation for $\beps$, and substituting the result 
into equation~(\ref{lvkram}), the following codeformational integral 
expansion is obtained, 
\begin{equation} 
\btau^p (t)  = - \, \int_{- \infty}^t d\!s \,  
n k_B T \, \tilde{G}(t-s)\, {\bgam}_{[1]}(t,s) 
\label{lvmemexp} 
\end{equation} 
where ${\bgam}_{[1]}$ is the 
codeformational rate-of-strain tensor~\cite{bird1} and the memory 
function $\tilde{G}(t)$ is given by \begin{equation} \tilde{G}(t)= 
{\cal H} \, e^{- (t / \lambda_H  \tau^*)  } \label{lvmemfn} 
\end{equation} The product $\lambda_H {\tau^*}$ is usually interpreted 
as a relaxation time, and ${\cal H}$ as a relaxation weight. Clearly, 
the incorporation of excluded volume effects increases the relaxation time 
in a good solvent relative to a theta solvent by a factor of $\tau^*$. 
 
The memory function $\tilde{G}(t)$ can now be used to derive the linear 
viscoelastic material properties.  Substituting $ G (t) = n k_B T 
\tilde{G}(t)$, into equation~(\ref{usf6}), leads to, \begin{equation} 
{\eta^\prime (\omega) \over \lambda_H n k_B T} = {\tau^* \, {\cal H} 
\over 1 + (\lambda_H \tau^* \, \omega)^2} \quad;\quad 
{\eta^{\prime\prime} (\omega) \over \lambda_H^2  nk_B T } = {\omega \, 
{\tau^*}^2 \, {\cal H} \over 1 + (\lambda_H \tau^* \,  \omega)^2} 
\label{etaprime} \end{equation} Upon taking the limit of $\omega \to 0$ 
we obtain, \begin{equation} {\eta_{p,0} \over \lambda_H n k_B T }= 
\tau^* \, {\cal H } \quad; \quad {\Psi_{1,0} \over \lambda_H^2 n k_B T 
} = 2 \, {\tau^*}^2 \, {\cal H} \label{lvetap0} \end{equation} 
 
At moderate to large values of the shear rate, it is not possible to 
obtain analytical expressions for the shear rate dependence of the  
viscometric functions, and consequently a numerical procedure is   
required. Since the second moment $\moms$ shares the symmetry of 
the flow field in simple shear flow, its Cartesian components can be 
denoted by,  
\begin{equation}  
\moms=\pmatrix{ s_1 & s_4 & 0 \cr 
\noalign{\vskip3pt} s_4 & s_2 & 0 \cr \noalign{\vskip3pt} 0 & 0 & s_3 
\cr }  
\label{momscompsf}  
\end{equation}  
Upon substituting equation~(\ref{momscompsf}) into 
equation~(\ref{gasecmom}), a system of four first order ordinary 
differential equations for the quantities $s_j, \, j=1, \ldots, 4$ is 
obtained.  Steady state viscometric functions, as functions of shear 
rate, can then be found by numerically integrating these equations with 
respect to time (using a simple Euler scheme) until steady state is 
reached. Results obtained by this procedure are discussed in 
section~7. 

\subsection{First order perturbation expansion in $z$} 

The influence of excluded volume effects on the universal shear rate 
dependence of viscometric functions has 
been studied, as mentioned earlier, by using renormalisation group 
methods~\cite{ottrg, zylkarg}. The renormalisation group theory 
approach is essentially a method for refining the results of a 
low order perturbation expansion in $z$, by introducing higher order 
interactions effects so as to remove the ambiguous definition of 
the bead size. Results of this approach, based on a  
$\delta$-function excluded volume potential, indicate that the 
presence of excluded volume has a non-trivial influence on the 
shear rate dependence of the viscometric functions. We have seen earlier 
in this work that as far as equlibrium swelling and zero shear 
rate properties are 
concerned, the use of a $\delta$-function excluded volume potential
leads to trivial results. In this section, a first order perturbation 
expansion in $z$---with a narrow Gaussian excluded 
volume potential---is constructed, in order to compare its predictions 
of shear rate dependence with those obtained with Brownian dynamics 
simulations, and with the Gaussian approximation. The dependence of 
the predictions on the width parameter $\mu$ is of particular interest.   

A first order perturbation expansion can be constructed, following the 
procedure suggested in references~\cite{ottrabrg, ottrg}, from the 
second moment equation~(\ref{nondimsecmom}). We assume that the 
configurational distribution function $\psi$ can be written as 
$\psi_\theta + \psi_z$, where $\psi_\theta$ is the distribution 
function in the absence of excluded volume, i.e. in a 
$\theta$-solvent, and $\psi_z$ is the correction to first order 
in the strength of the excluded volume interaction. The averages 
performed with these contributions will be denoted 
by $\avel \cdots \aver_\theta$ and $\avel \cdots \aver_z$, respectively.

On equating terms of equal order, equation~(\ref{nondimsecmom}) can 
be rewritten as two equations, namely, a zeroth order second moment 
equation and a first order second 
moment equation. The zeroth order equation---which is linear 
in the moment $\avel \bQ^* \bQ^* \aver_\theta$---is the well 
known second moment equation for Hookean dumbbells in a 
$\theta$-solvent~\cite{bird2}. It has the analytical 
solution~\cite{bird2},
\begin{equation}
\avel \bQ^* \bQ^* \aver_\theta  = 
\bu - \, \int_{- \infty}^{t^*} d\!s^* \,  
e^{-(t^*-s^*) } \, {\bgam}_{[0]}(t^*, s^*) 
\label{qqtheta}
\end{equation}
where, ${\bgam}_{[0]}$ is the codeformational relative strain 
tensor~\cite{bird1}. The first order second moment equation 
has the form, 
\begin{equation} 
{d \over dt^*} \avel \bQ^* \bQ^* \aver_z = 
\bk^* \cdot \avel \bQ^* \bQ^* \aver_z 
+ \avel \bQ^* \bQ^* \aver_z \cdot {\bk^*}^T - 
\avel \bQ^* \bQ^* \aver_z + {z \over \mu^5} \, 
\avel  e^{- ({Q^*}^2 / 2 \, \mu^2) } \, \bQ^* \bQ^* \aver_\theta 
\label{secmomz} 
\end{equation} 
The $\theta$-solvent distribution function $\psi_\theta$ is a 
Gaussian~\cite{bird2}, and consequently, the complex moment 
on the right hand side of equation~(\ref{secmomz}) can be 
reduced to a function of $ \avel \bQ^* \bQ^* \aver_\theta$. 
The following equation is obtained on performing this reduction, 
\begin{equation}  
{d \,  \over dt^* } \avel \bQ^* \bQ^* \aver_z  =  
\bk^* \cdot \avel \bQ^* \bQ^* \aver_z + \avel \bQ^* \bQ^* \aver_z
\cdot {\bk^*}^T - \avel \bQ^* \bQ^* \aver_z + \bY 
\label{redsecmomz}  
\end{equation} 
where, 
$$ \bY = {z \over {{\sqrt {{\rm det} \,  
\lbrack \, {\avel \bQ^* \bQ^* \aver_\theta} + \, 
\mu^2 \, {\bu } \rbrack }}}} \, \bigl\lbrack 
\avel \bQ^* \bQ^* \aver_\theta + \mu^2 \, {\bu} \, \bigr\rbrack^{\, -1} 
\cdot  \avel \bQ^* \bQ^* \aver_\theta  $$

Clearly, equation~(\ref{redsecmomz}) could have also been derived 
by expanding the second moment equation for the Gaussian 
approximation~(\ref{gasecmom}) to first order in $z$. 
It follows, therefore, that the Gaussian 
approximation is exact to first order in $z$. 
This is also the situation in the case of the Gaussian approximation 
introduced for the treatment of hydrodynamic interaction effects, 
where it was found to be exact to first order in the
strength of hydrodynamic interaction, $h^*$~\cite{ottrabrg}.

Equation~(\ref{redsecmomz}) is a system of linear inhomogeneous ordinary
differential equations, whose solution is,
\begin{equation}
\avel \bQ^* \bQ^* \aver_z  = \int_{- \infty}^{t^*} d\!s^* \,  
e^{-(t^*-s^*) } \, \bE(t^*, s^*) \cdot \bY \cdot  \bE{^T}(t^*, s^*) 
\label{qqz}
\end{equation}
where, $\bE$ is the displacement gradient tensor~\cite{bird1}. 

The expression for the stress tensor~(\ref{nondimkram}) can also be 
expanded to first order in $z$. After reduction of complex moments 
to second moments, the stress tensor depends only on 
the second moments $\avel \bQ^* \bQ^* \aver_\theta $ and 
$\avel \bQ^* \bQ^* \aver_z$. Equations~(\ref{qqtheta}) and~(\ref{qqz}) 
may then be used to derive the following first order perturbation theory
expression for the stress tensor in arbitrary homogeneous flows,
\begin{equation}
{\btau^p \over nk_BT} = \bY + \int_{- \infty}^{t^*} d\!s^* \,  
e^{-(t^*-s^*) } \, \left( {\bgam}_{[0]}(t^*, s^*) 
-  \bE(t^*, s^*) \cdot \bY \cdot  \bE{^T}(t^*, s^*) \right) 
\label{pertau}
\end{equation}
Note that $\bY$, the direct contribution to the stress tensor,  
is isotropic only in the limit $\mu \to 0$. 

The form in steady shear flow, of the tensors ${\bgam}_{[0]}$ and $\bE$,   
has been tabulated in reference~\cite{bird1}. Using the expression for
the stress tensor~(\ref{pertau}), and the definition of the viscometric 
functions~(\ref{sfvis}), the following first order perturbation theory 
results for the viscometric functions are obtained,
\begin{eqnarray} 
{\eta_{p} \over \lambda_H n k_{\rm B} T } &=&  1 +
\left( {1 + \mu^2 + \lambda_H^2 {\dot \gamma}^2 \over 
\sqrt{1 + \mu^2 } \, \Delta^{3/2} } \right) \, z
\label{etaper} \\ 
{\Psi_{1} \over \lambda_H^2 n k_{\rm B} T }  &=& 
2  + 2 \left( {1 + 2 \, \mu^2 + \lambda_H^2 {\dot \gamma}^2 \over 
\sqrt{1 + \mu^2 } \, \Delta^{3/2} } \right) \, z 
\label{Psi1per}  
\end{eqnarray}  
where, $\Delta=(1 + \mu^2)[1 + \mu^2 + 2 \lambda_H^2 
{\dot \gamma}^2] - \lambda_H^2 {\dot \gamma}^2$. In the limit 
of $\lambda_H {\dot \gamma}$ going to zero, the expression 
for the viscosity~(\ref{etaper}) reduces to the expression derived 
earlier in section~5.1---using the retarded motion expansion 
and the equilibrium perturbation expansion---for the zero shear rate 
viscosity. One can also show that ${\rm tr}  \avel \bQ^* \bQ^* 
\aver $ reduces to the equilibrium moment $q_1$ [see 
equation~(\ref{equimom})], in the limit $\lambda_H {\dot \gamma} 
\to 0$. 

The first order perturbation results are compared with 
simulation results, and with results of the Guassian 
approximation, in section~7. 

\section{Fixman's Theory} 
 
Many years ago, in path-breaking seminal work, Fixman~\cite{fix} 
considered the simultaneous inclusion of hydrodynamic interaction and 
excluded volume in bead-spring chain models for dilute polymer 
solutions. In order to render the problem solvable, Fixman introduced a 
number of approximations. Since we are only concerned with excluded 
volume in the context of Hookean dumbbells in this work, we shall only 
consider those approximations which are relevant in this context.  The 
introduction of the quadratic potential, the governing equations of 
Fixman's theory, and the calculation of material functions predicted by 
the theory in simple shear flow are considered in this section. 
 
\subsection{The quadratic potential} 
With regard to excluded volume, the most crucial approximation of
Fixman~\cite{fix} is the replacement of the delta function potential with
a quadratic potential.  By adopting a Boson operator formulation of the
governing equations, Fixman has shown that the delta 
function potential~(\ref{delpot}) may be represented by, 
\begin{equation}
E^* \, (\bQ^*) = {1 \over 2}\, \bQ^* \cdot {\bGF^* } \cdot \bQ^*  
\label{bosonE}
\end{equation}
where ${\bGF^*}$ is a symmetric function of various configuration
dependent quantities introduced in the Boson operator formalism.  From
this expression it is clear that a quadratic potential for the excluded
volume may be obtained by replacing the fluctuating quantity $\bGF^*$
with an average. Fixman obtains a quadratic potential
by replacing $\bGF^*$ with a configuration dependent average, as described 
below.

As a result of replacing $\bGF^*$ with its average, one can show 
from equation~(\ref{bosonE}) that, 
\begin{equation}
\avel \, \bGF^* \, \aver = \avel \, {\partial \over \partial \bQ^*}
\,{\partial E^* \over \partial \bQ^*}\, \aver 
\label{avgG}
\end{equation}
In other words, for any given potential $E^*$, one can find 
$\avel \, \bGF^* \, \aver $ 
provided that the non-equilibrium distribution function $\psi (\bQ,t)$ 
with which to carry out the average on the right hand side of 
equation~(\ref{avgG}) is known.
It turns out that $\psi (\bQ,t)$, in
the presence of a quadratic excluded volume potential and with
consistently averaged hydrodynamic interaction~\cite{ottca1, prakbk}, 
is a Gaussian distribution. Evaluation of the average 
for a $\delta$--potential~(\ref{delpot}), with a Gaussian 
distribution~(\ref{gausdist}), leads to the following 
non-dimensional quadratic potential,  
\begin{equation}
E^* \,(\bQ^*)=-  {1 \over 2} \, { z \over {\sqrt{ {\rm det}
\moms}}} \,\, \bQ^* \cdot \moms^{-1} \cdot \bQ^* 
\label{modquadpot}
\end{equation}
It is straightforward to show that the potential~(\ref{modquadpot}) 
leads to an unphysical {\it non-central} excluded volume force 
between the beads. Fixman, perhaps for this reason, introduces a 
further approximation which consists of replacing the above 
potential with the following simpler form,
\begin{equation}
E^* \,(\bQ^*)=- {1 \over 2} \, {z \over \alpha^2} 
\, {{Q^*}^2 \over \, {\sqrt{ {\rm det} \moms}}} 
\label{quadpot}
\end{equation}
where, $\alpha$ is defined as before by  equation~(\ref{alpha}). 
However, in this case, $\alpha$ obeys the consistency relation, 
\begin{equation} 
z = {(\, \alpha^2 - 1 )\, \alpha^3 } 
\label{fixalp} 
\end{equation} 
Interestingly enough, equation~(\ref{equirel}) reduces to 
equation~(\ref{fixalp}) in the limit $\mu \to 0$ (which corresponds 
to a $\delta$--potential). 

It is appropriate here to note
that while Fixman has presented all his arguments for bead-spring chain
models, the form of Fixman's potential for dumbbells given above can be
found in the book by Larson~\cite{larson}.

The consequences of adopting the quadratic excluded volume 
potential~(\ref{quadpot}) are briefly discussed in the 
following section. It is worthwhile to 
point out here that Fixman's original formulation of the problem 
was not in terms of the Gaussian distributions and second moments 
discussed below. Rather, his attempt was to
directly solve the diffusion equation~(\ref{diff}) for the configurational
distribution function, and then carry out the average in 
equation~(\ref{kram}) to obtain
the rheological properties. It is, however, possible to discuss his
approach within the framework developed subsequently by
{\" O}ttinger~\cite{ottca1} and this is the procedure that 
is adopted here.  A graphic exposition of Fixman's algorithm is 
given in reference~\cite{larson}.

\subsection{The governing equations} 
 
With excluded volume interactions described by the quadratic 
potential~(\ref{quadpot}), the diffusion equation~(\ref{diff}) becomes 
linear in the bead-connector vector. As a result, the diffusion 
equation is exactly satisfied by a Gaussian 
distribution~(\ref{gausdist}). In Fixman's theory therefore, a 
tractable model is obtained not by approximating the distribution 
function, as in the case of the Gaussian approximation, 
but by introducing the quadratic potential~(\ref{quadpot}). 
 
While $\psi (\bQ,t)$ is a Gaussian both in the Gaussian approximation 
and in Fixman's theory, the second moment which completely determines 
these distributions is different in the two cases. In Fixman's theory, 
the non-dimensional second moment $\moms$ is governed by the equation, 
\begin{equation} 
{d \over dt^*} \avel \bQ^* \bQ^* \aver = \bk^* \cdot 
\avel \bQ^* \bQ^* \aver + \avel \bQ^* \bQ^* \aver \cdot {\bk^*}^T - 
\left[ \, 1 - {z \over \alpha^2 {\sqrt{ {\rm 
det} \avel \bQ^* \bQ^* \aver}}} \, \right] \, \avel \bQ^* \bQ^* \aver + \bu 
\label{fixsecmom} 
\end{equation} 
Note that the second moment equation reduces to equation~(\ref{fixalp}) 
at equilibrium.
 
The Giesekus expression for the stress tensor has 
the form, \begin{equation} {\btau^p \over n k_B T } =  
-  \left[ \, 1 - {z \over \alpha^2 {\sqrt{ {\rm 
det} \avel \bQ^* \bQ^* \aver}}} \, \right] \, \moms + 
\bu \label{fixkram} \end{equation} As a result, the stress tensor in 
Fixman's theory, for any flow situation, may be obtained once 
equation~(\ref{fixsecmom}) is solved for $\moms$. 
 
It is clear from  equation~(\ref{fixkram}) that in Fixman's theory, as 
for $\mu > 0$ in the Gaussian approximation, rheological properties are 
affected both directly and indirectly by the presence of excluded 
volume. 
 
As will be shown in the section below, at {\it steady state} in simple 
shear flow, the problem of solving the governing equation for the second 
moments~(\ref{fixsecmom}), reduces to one of solving a single nonlinear 
algebraic equation.  In 
the linear viscoelastic limit, however, analytical expressions for the 
various properties can be derived in the same manner as described 
earlier for the Gaussian approximation.  Indeed, it can be shown that 
the linear viscoelastic properties are given by 
equations~(\ref{etaprime}) and~(\ref{lvetap0}), where the quantities 
${\cal H}$ and $\tau^* $ are now given by, \begin{equation} {\cal H} = 
1 \quad;\quad \tau^* = \alpha^2 \label{fixlvspk} \end{equation} 
 
In steady simple shear flow, substituting equation~(\ref{momscompsf}) 
for $\moms$ and equation~(\ref{ssf1}) for $\bk$, into the second moment 
equation~(\ref{fixsecmom}), leads to the following equations for the 
components of $\moms$, $$s_1 = ( 1 + 2 \, \lambda_H^2 {\dot \gamma}^2 
\, s_2^2 \, ) \, s_2 \quad;\quad s_3 = s_2 \quad;\quad s_4 = \lambda_H 
\, {\dot \gamma} \, s_2^2 $$ where, $s_2$ must satisfy the nonlinear 
algebraic equation, \begin{equation} s_2^{3 / 2} -  s_2^{1 / 2} = 
{\alpha(\alpha^2 -1) \over \sqrt{ 1 + \lambda_H^2 {\dot \gamma}^2 \, 
s_2^2 } } \label{sfs2} \end{equation} 
 
The normalised viscometric functions can be found by using the 
definitions~(\ref{sfvis}), and the results above for the zero shear 
rate properties, \begin{equation} {\eta_p \over \eta_{p,0}} = {s_2 
\over \alpha^2} \quad;\quad {\Psi_1 \over \Psi_{1,0}} = {s_2^2 \over 
\alpha^4} \quad;\quad \Psi_2 = 0 \label{fixetar} \end{equation} 

The nonlinear algebraic equation~(\ref{sfs2}) is solved here with a
Newton-Raphson scheme. The material functions predicted by Fixman's
theory are compared with the predictions of the narrow Gaussian
potential in the section below.

\section{Results and Discussion}

The predictions of rheological properties in simple shear flow, by the
various theories for the excluded volume effect, are compared in this
section.  Predictions in the limit of zero shear rate are first
considered below, and those at finite non-zero shear rates
subsequently.

\subsection{Zero shear rate properties}

\begin{figure}[!htb] \centerline{ \epsfxsize=4in \epsfbox{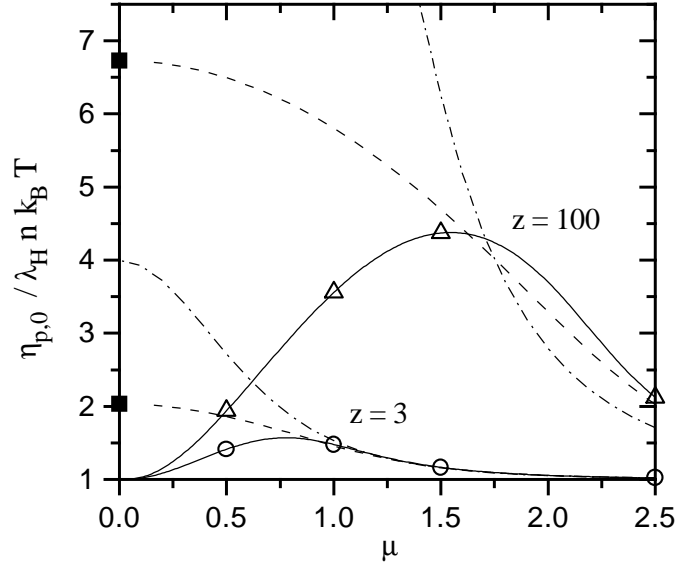}}
\caption{ \footnotesize Non-dimensional zero shear rate viscosity
versus the extent of excluded volume interaction $\mu$, for two values
of the strength of the interaction $z$.  The continuous lines are exact
predictions obtained by numerical quadrature, 
the triangles and circles are results of 
Brownian dynamics simulations, the dashed and the dot-dashed lines 
are the approximate predictions of the Gaussian approximation, and the 
first order perturbation theory, respectively, and the filled squares 
are the predictions of Fixman's theory.  The error bars in the Brownian 
dynamics simulations cannot be resolved within the line thickness.}
\label{fig1} \end{figure}

Figure~{\ref{fig1}} is a plot of $(\eta_{p,0}/ \lambda_H \, nk_BT) $
versus $\mu$ for $z=3$ and $z=100$. The continuous curves are exact
predictions obtained by numerical quadrature, while the triangles and
circles are exact results of Brownian dynamics simulations 
carried out at equilibrium (without variance reduction) 
[see equation~(\ref{etap02})].  The dashed lines are the
predictions of the Gaussian approximation for the narrow Gaussian
potential [{\em i.e}.\ equation~(\ref{lvetap0}), with $\tau^*$ and
${\cal H}$ given by equations~(\ref{lvtau}) and~(\ref{lvH}),
respectively], the dot-dashed curve is the prediction of the 
first order perturbation theory [{\em i.e}.\ equation~(\ref{etaper}) in 
the limit $\lambda_H {\dot \gamma} \to 0$], and the filled squares 
are the predictions of Fixman's theory [{\em i.e}.\ 
equation~(\ref{lvetap0}), with $\tau^*$ and ${\cal H}$ given by 
equations~(\ref{fixlvspk})]. The parameter $\mu$ does not
enter into Fixman's theory, however, these values are plotted
corresponding to $\mu=0$, since the quadratic potential is used in
Fixman's theory as an approximation for the $\delta$-potential.

The first feature to be noticed in figure~{\ref{fig1}} is the
reassuring closeness of the exact results obtained by using the
retarded motion expansion and by Brownian dynamics simulations. The
retarded motion expansion provides a means of validating the results of
Brownian dynamics simulations.

In the limit $\mu \to 0$, and for large values of $\mu$, the continuous
curves and Brownian dynamics simulations reveal that, as expected, the
exact predictions of the narrow Gaussian potential tend to the
$z$-independent $\theta$-solvent value, $(\eta_{p,0}/ \lambda_H \, nk_BT)
= 1$.  This implies, as pointed out earlier, that the use of a
$\delta$-function potential to represent excluded volume interactions
does not lead to any change in the zero shear rate viscosity
prediction.  On the other hand, figure~{\ref{fig1}} seems to suggest
that a finite range of excluded volume interaction is required to cause
a change from the $\theta$-solvent value.  Away from these limits, at
non-zero values of $\mu$, the narrow Gaussian potential predicts an
increase in the value of zero shear rate viscosity. The existence of
shear thinning in good solvents can be attributed to this increase.
This follows from the fact that at high shear rates, as the effect of
the excluded volume interaction diminishes, the viscosity is expected
to return to its $\theta$-solvent value. We shall see later that this
expectation is indeed justified.

\begin{figure}[!htb] \centerline{ \epsfxsize=4in \epsfbox{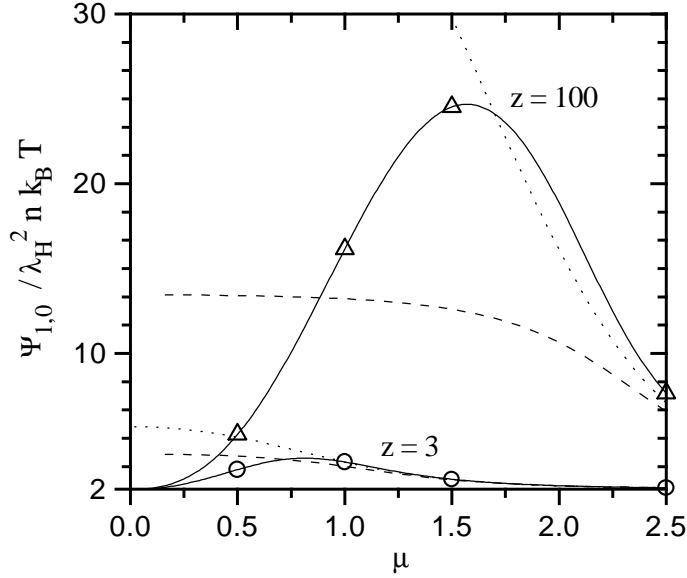}}
\caption{ \footnotesize Non-dimensional zero shear rate first normal
stress difference versus $\mu$, for two values of $z$.  The continuous
lines are exact predictions obtained by numerical quadrature, 
the triangles and circles are results of Brownian dynamics
simulations, the dashed lines are approximate predictions using the
Gaussian approximation, and the dotted lines are approximate
predictions using the uniform expansion model. The error bars 
in the Brownian dynamics simulations cannot be resolved within 
the line thickness. } \label{fig2}
\end{figure}

The dashed lines in figure~{\ref{fig1}} indicate that in the limit of
zero shear rate, for a given value of $z$, the Gaussian approximation
is reasonably accurate above a certain value of $\mu$. This limiting
value of $\mu$ appears to be smaller for smaller values of $z$.  A
similar behavior is also observed with regard to the prediction of the
zero shear rate first normal stress difference (see
figure~{\ref{fig2}}).  Thus it appears that the exact configurational
distribution function $\psi (\bQ, t)$ becomes increasingly non-Gaussian
as the narrow Gaussian potential becomes narrower, and as the strength
of the excluded volume interaction becomes larger.

For the large values of $z$ considered in figure~{\ref{fig1}}, 
results obtained with the first order perturbation expansion in $z$ 
cannot be expected to be accurate. It is clear from the dot-dashed lines   
that the perturbation expansion results deviate significantly 
from exact results for small values of $\mu$. However, they become 
increasingly accurate as $\mu$ increases, for a given value of $z$. 
This can be understood 
by considering equation~(\ref{etaper}), which indicates that in 
the limit $\lambda_H {\dot \gamma} \to 0$, the 
first order correction to the $\theta$-solvent value increases as
$z$ increases, but decreases as $\mu$ increases. 

As $\mu$ increases from zero, values of the zero shear rate viscosity 
predicted by the Gaussian approximation, approach the exact values 
more rapidly than the predictions of the first order perturbation 
theory, for a given value of $z$. The Gaussian approximation is a 
non-perturbative approximation---however, it was shown in section~5.4, 
to be exact to first order in $z$. One way to understand this is  
to consider the Gaussian approximation to consist of an infinite 
number of higher order terms, whose nature is unknown. In this sense, 
it is an {\em uncontrolled} approximation, which remains accurate at values 
of $z$ and $\mu$, where the first order perturbation expansion becomes  
inaccurate. As will be seen shortly, these remarks apply also to the 
results obtained at finite shear rate.   

The difference in the prediction of the zero shear rate viscosity by
Fixman's theory and by the narrow Gaussian potential is evident in
figure~{\ref{fig1}}. It is also worth noting that, although the
relaxation weight and the relaxation time are different in Fixman's
theory and in the Gaussian approximation for $\mu=0$, they lead to the
same prediction of the zero shear rate viscosity. This is, however, not
true for the zero shear rate first normal stress difference.  The ratio
$U_{\Psi \eta}$, defined by~\cite{ottbk}, 
\begin{equation} 
U_{\Psi \eta} =  {n k_B T \Psi_{1} \over \eta_{p}^2 }
\label{upsieta}
\end{equation}  
is equal, in the zero shear rate limit, to $(2 / \alpha^2)$ in the 
Gaussian approximation, while it has a constant value of 2 in 
Fixman's theory.

Both the uniform expansion model and the Gaussian approximation use
Gaussian distributions in order to evaluate averages. However, the
uniform expansion model uses different Gaussian distributions for
different equilibrium averages, such that the best approximation is
obtained. While this does not lead to any difference in the prediction
of the zero shear rate viscosity by the two approximations,
figure~{\ref{fig2}} reveals that, at small enough values of $\mu$,
there is a significant difference in the prediction of the zero shear
rate first normal stress difference. Clearly, the uniform expansion
model continues to be a reasonable approximation for values of $\mu$ at
which the Gaussian approximation is no longer accurate. However, even
the uniform expansion model leads to a poor approximation at
sufficiently small values of $\mu$.

\begin{figure}[!htb] 
\centerline{\epsfxsize=4in \epsfbox{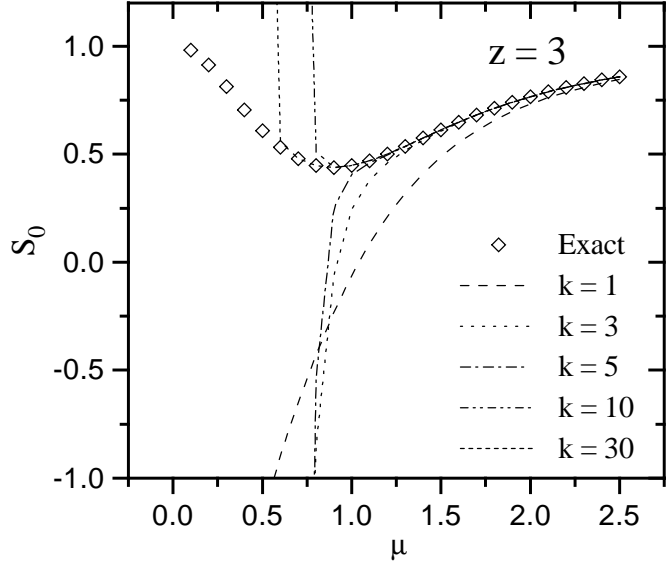}}
\caption{ \footnotesize
The sum $S_0$ versus $\mu$, for different numbers of summed terms 
$k$ in the perturbation expansion [see equation~(\ref{s0s1})]. 
The exact results are obtained by numerical quadrature.} 
\label{fig3}
\end{figure}

In the discussion of the equilibrium perturbation expansion in 
section~5.1.2, it was pointed out that, in principle, the 
equilibrium moment $q_1$ can be obtained for any non-zero value 
of $\mu$, provided that enough numbers of terms in the series for the 
quantities $S_0$ and $S_1$, are summed [see equations~(\ref{s0s1})]. 
Figure~3 displays the sum $S_0$, for different numbers of summed 
terms in the perturbation expansion, as a function
of $\mu$, for $z = 3.0$.  Exact results are obtained by noting
that $S_0 = \sqrt{2 / \pi} \, I_0$, and evaluating $I_0$ by numerical
quadrature. Clearly, more terms of the expansion are required 
for convergence as $\mu$ decreases.  The terms
of the series, which are alternating in sign, keep increasing rapidly in
magnitude, until $n$ is approximately $ > (z/\mu^3) $, before they 
begin to decrease. Therefore, as $z$ increases, or as $\mu$
decreases, more and more terms are required for the sum to converge.
Above a threshold value of $(z/\mu^3)$ however, it becomes impossible to
evaluate $S_0$ since the round-off errors due to the summation of large
numbers makes the perturbation expansion numerically useless. A 
similar problem is also encountered while evaluating the sum $S_1$.
In short, the hope of extrapolating finite $\mu$ results to the limit $\mu=0$,
cannot be realised. Therefore, in the case of Hookean dumbbells, one
cannot obtain equilibrium moments for a delta function excluded volume
potential by using a narrow Gaussian potential and considering a
perturbation expansion in the limit $\mu \to 0$.

\subsection{Steady state viscometric functions}

\begin{figure}[!htb] \centerline{ \epsfxsize=4in \epsfbox{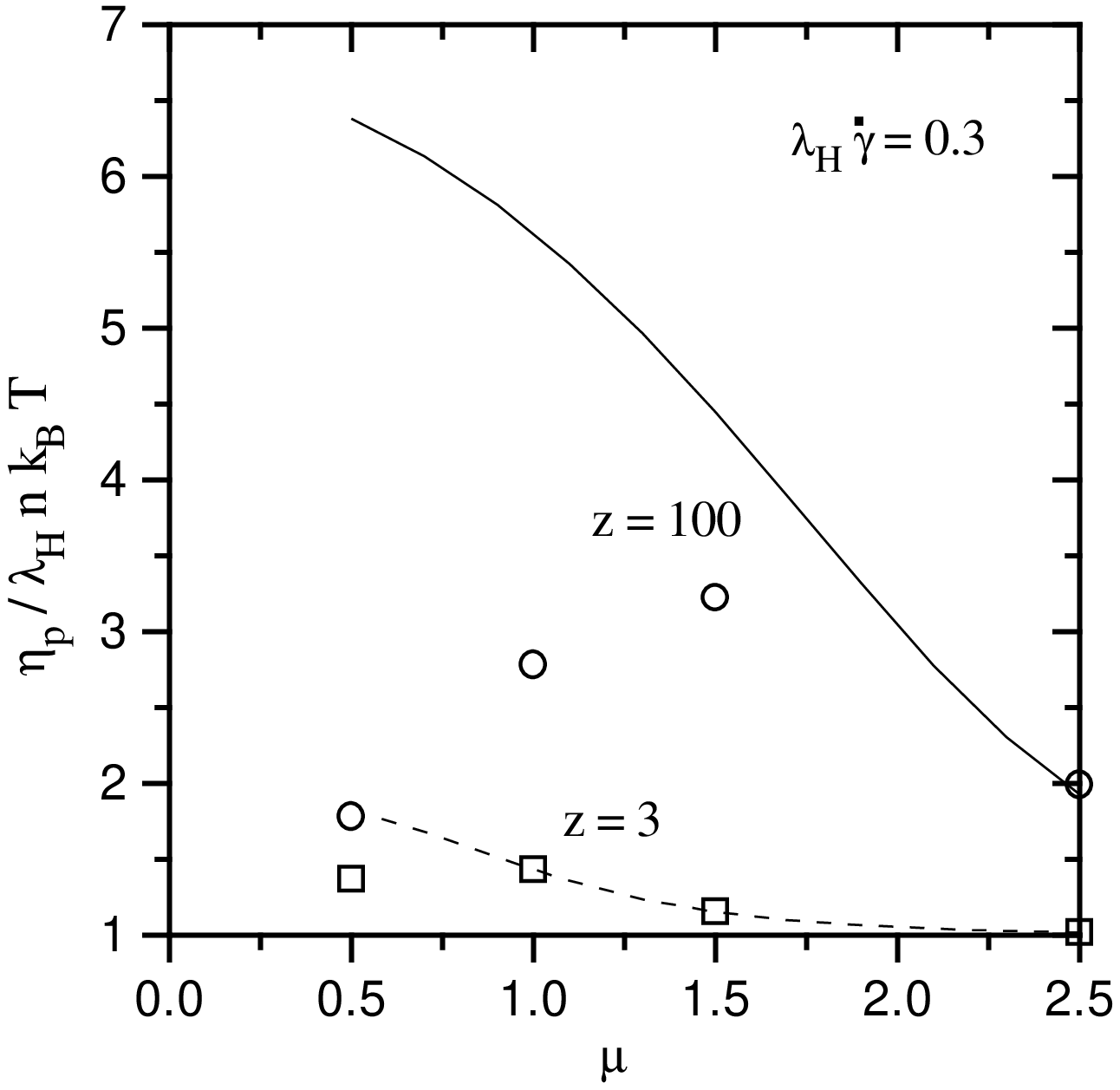}}
\centerline{ \epsfxsize=4in \epsfbox{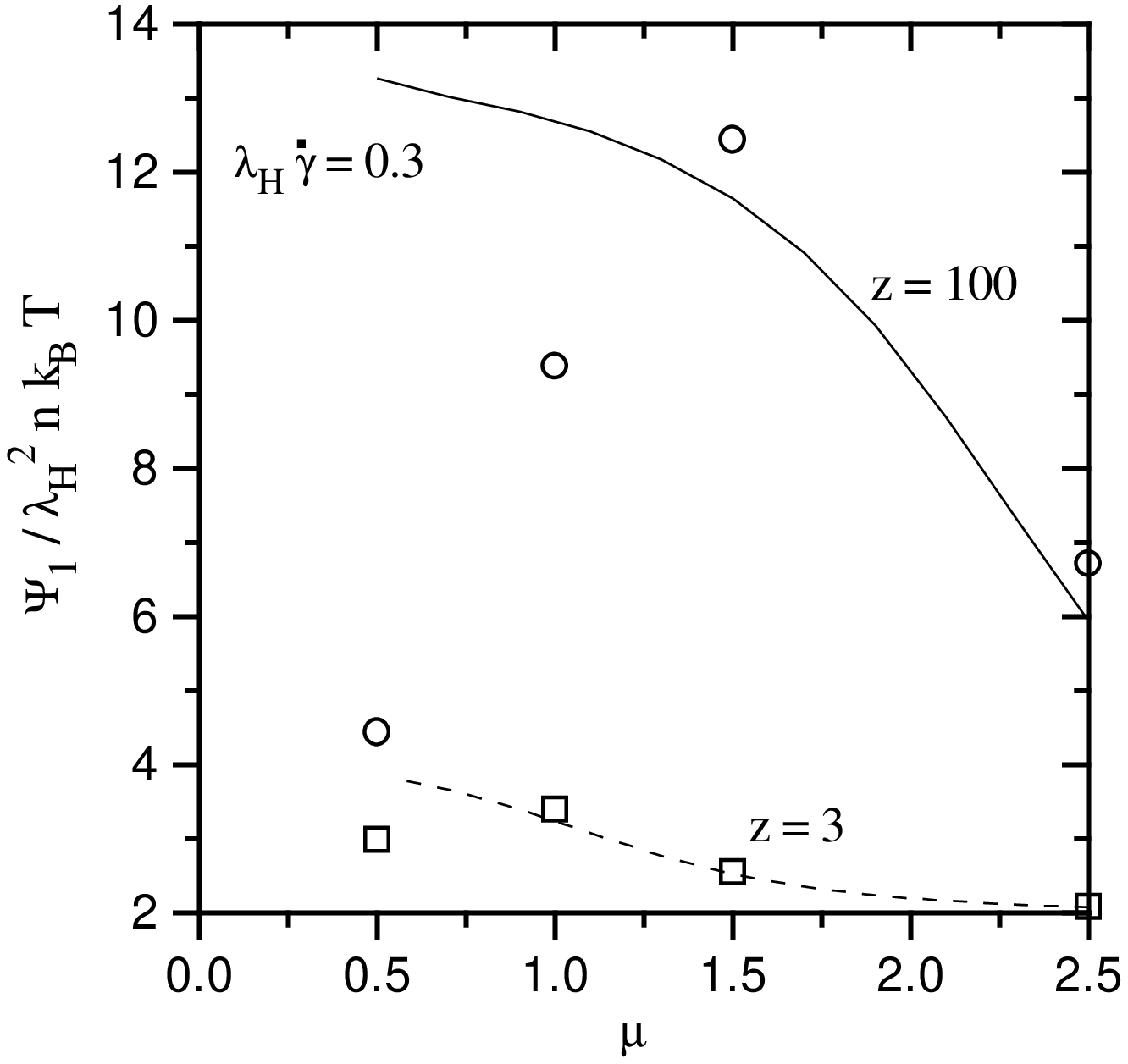}} \caption{\footnotesize
Non-dimensional viscosity and first normal stress difference versus
$\mu$ for two values of $z$, at a non-dimensional shear rate $\lambda_H
{\dot \gamma} = 0.3$.  The squares and circles are results of Brownian
dynamics simulations, and the dashed and  continuous lines are 
the predictions of the Gaussian approximation, for $z=3$ and $z=100$, 
respectively. The error bars 
in the Brownian dynamics simulations cannot be resolved within 
the line thickness.} \label{fig4}
\end{figure}

The results of Brownian dynamics simulations 
(without variance reduction) displayed in
figure~\ref{fig4} reveal that the dependence of the viscosity and the
first normal stress difference on $\mu$, at a value of the
non-dimensional shear rate $\lambda_H {\dot \gamma} = 0.3$, is similar
in shape to the dependence observed in the limit of zero shear rate. At
small and large values of $\mu$ the material functions tend to the
$\theta$-solvent value, and exhibit a maximum at some value in between.
Since, even at this non-zero value of shear rate, it appears that the
viscosity and the first normal stress difference remain at their 
$\theta$-solvent values for $\mu =0$, it implies that the use of a
$\delta$-potential to represent excluded volume interactions would not
predict any shear thinning. On the other hand, as we shall see
subsequently, a quadratic potential does predict substantial shear
thinning. At $\lambda_H {\dot \gamma} = 0.3$, the Gaussian
approximation seems to be accurate above roughly the same values of
$\mu$ as were observed in the limit of zero shear rate.

\begin{figure}[!htb] 
\centerline{\epsfxsize=4in \epsfbox{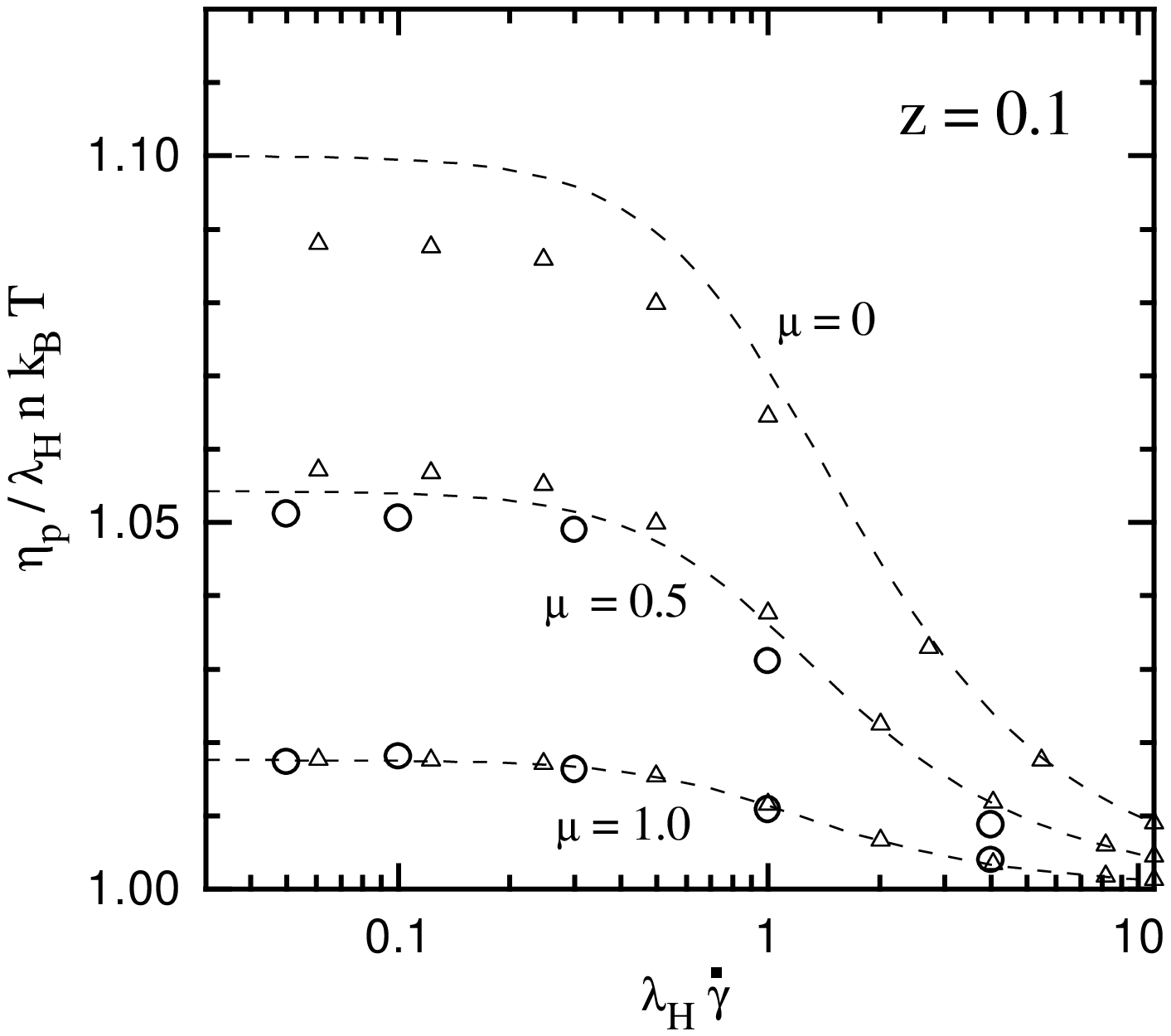}}
\centerline{ \epsfxsize=4in \epsfbox{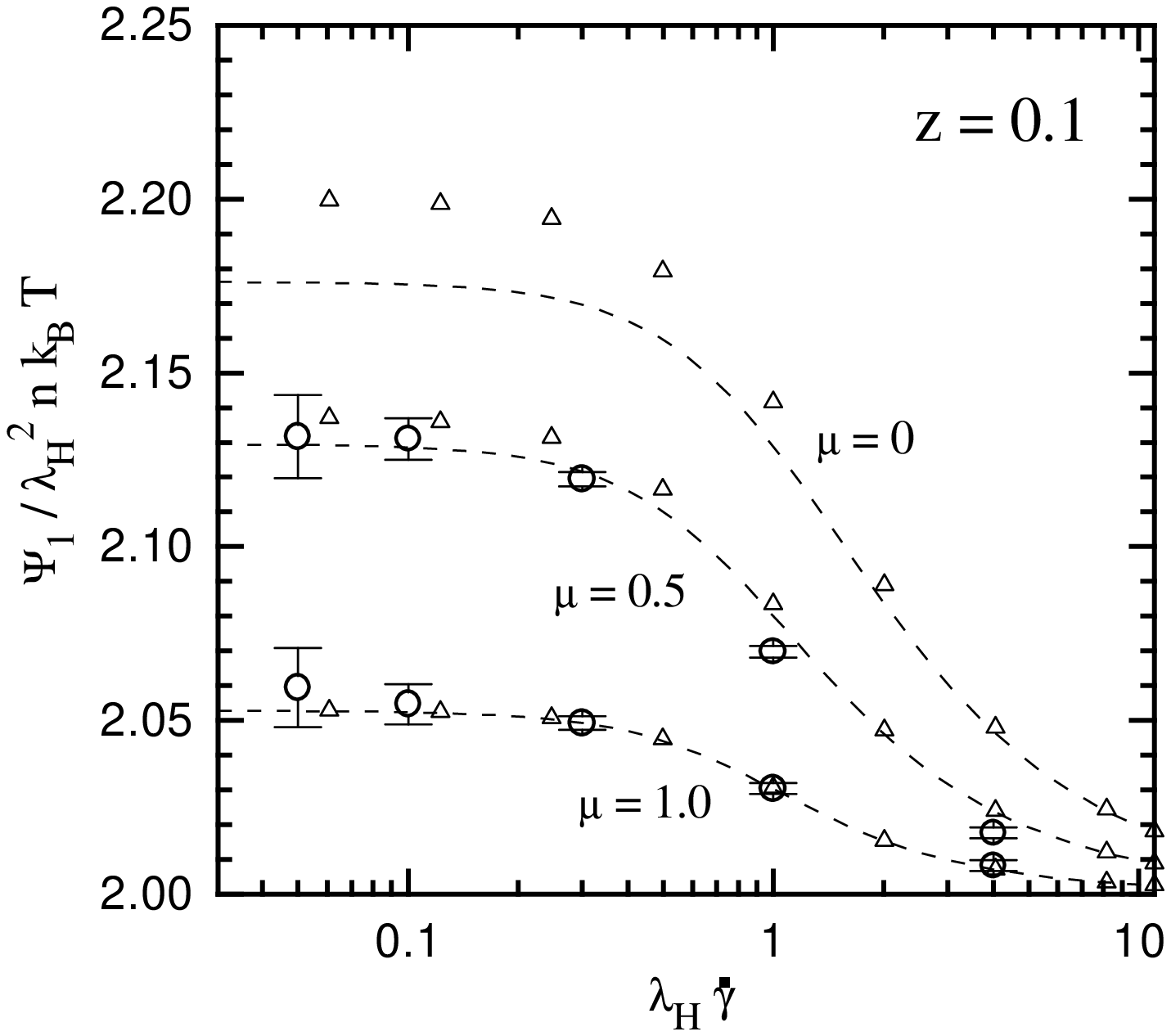}} 
\caption{ \footnotesize
Non-dimensional viscosity and first normal stress difference versus
non-dimensional shear rate $\lambda_H {\dot \gamma}$, 
for three values of $\mu$ at $z = 0.1$.
The circles are results of Brownian dynamics simulations, 
the dashed lines are the predictions of the Gaussian approximation, 
and the triangles are the predictions of the first order 
perturbation expansion. In the viscosity plot, the error bars in 
the Brownian dynamics simulations are smaller than the size of the 
symbols. }
\label{fig5} \end{figure}

\begin{figure}[!htb] 
\centerline{\epsfxsize=4in \epsfbox{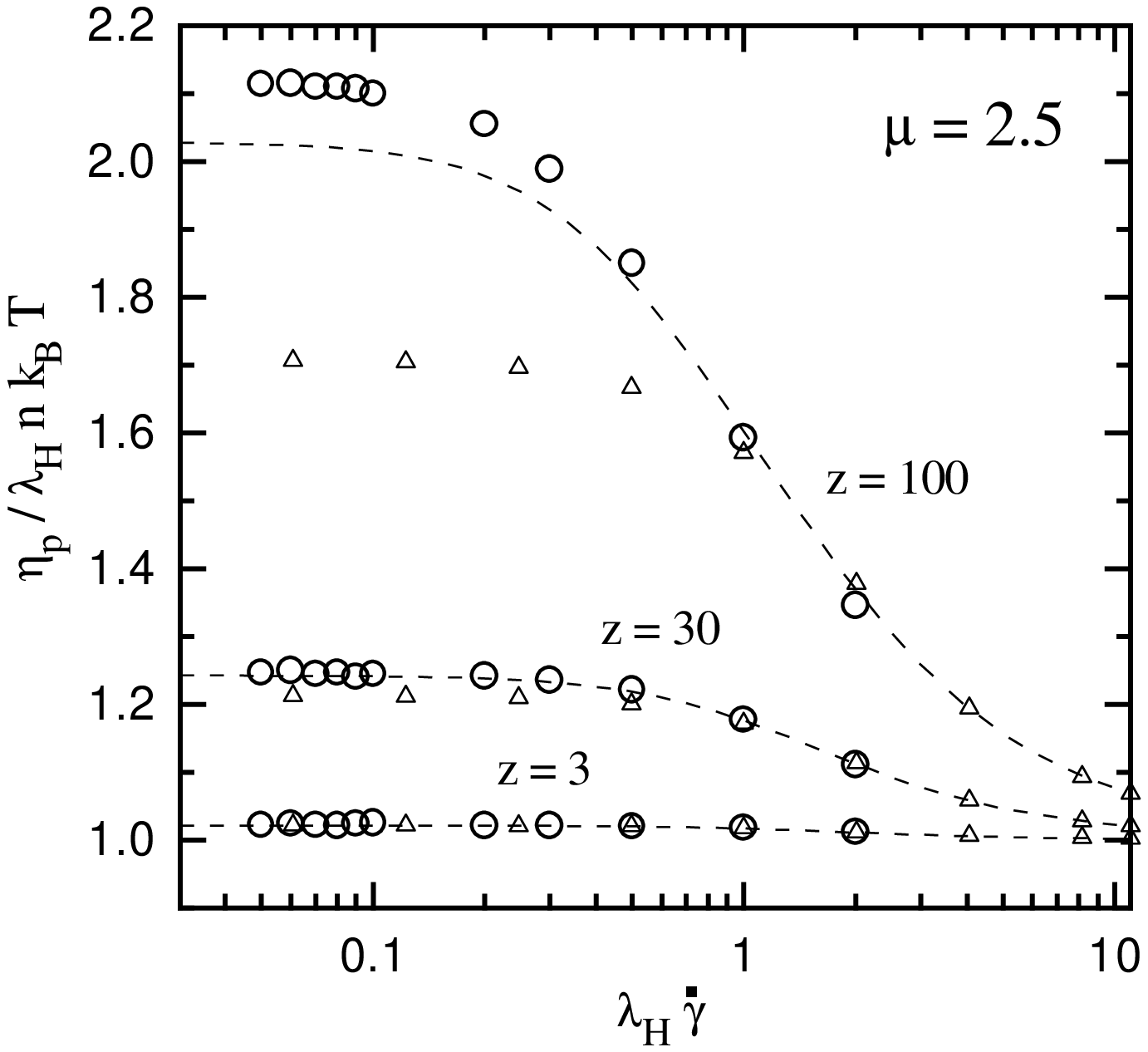}}
\centerline{ \epsfxsize=4in \epsfbox{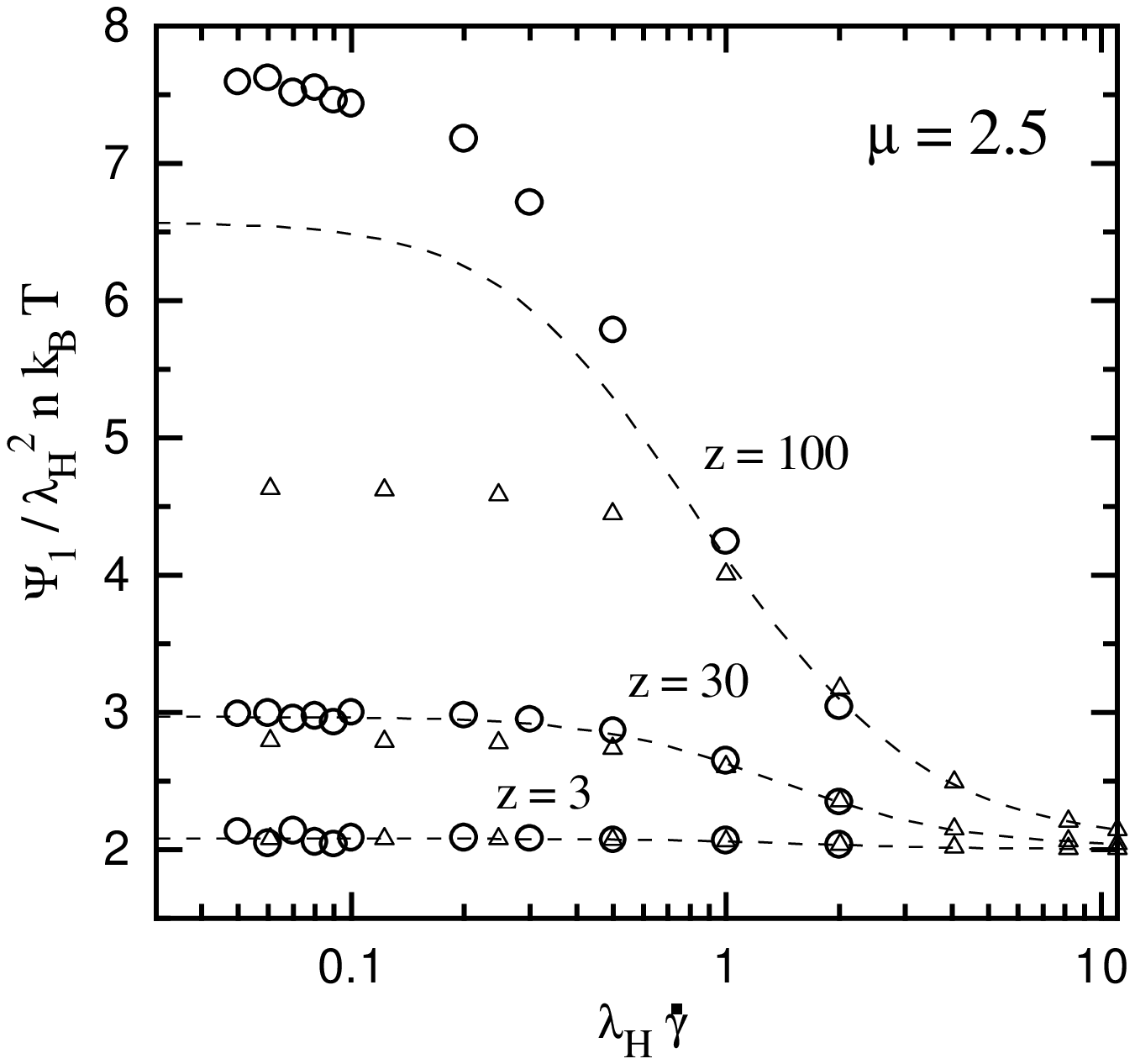}} 
\caption{ \footnotesize
Non-dimensional viscosity and first normal stress difference versus
$\lambda_H {\dot \gamma}$ for three values of $z$, at $\mu = 2.5$.
The symbols are as indicated in the caption to figure~\ref{fig5}. 
The error bars in the Brownian dynamics simulations are smaller 
than the size of the symbols.}
\label{fig6} \end{figure}

Figures~\ref{fig5} and~\ref{fig6} are plots of non-dimensional 
viscosity and first normal stress difference versus the non-dimensional 
shear rate $\lambda_H {\dot \gamma}$.  Figure~\ref{fig5} displays the 
dependence of these viscometric functions 
on the parameter $\mu$, for a fixed value of $z = 0.1$, 
while figure~\ref{fig6} displays the dependence on the parameter $z$,  
for a fixed value of $\mu=2.5$. The prediction of shear thinning for 
non-zero values of $\mu$ is apparent, and our earlier expectation 
in this direction is justified. In particular, the predictions of 
Brownian dynamics simulations, the Gaussian approximation, and the 
first order perturbation theory tend to $\theta$-solvent values at 
high shear rates.  

Shear thinning, which is seemingly physically meaningful, 
is also predicted, as can be seen from figure~\ref{fig5}, 
by both the Gaussian approximation and the first order perturbation 
theory, for $\mu=0$. This corresponds to a $\delta$-function excluded 
volume potential, and is clearly an artifact of the perturbation 
expansion, since rigorous calculations indicate a trivial result. It remains 
to be seen if the situation is different in the limit of long chains.  

For small enough values of $z$, and large enough values of 
$\mu$, the results of the Gaussian approximation, and the first order 
perturbation expansion, agree exceedingly well with the exact results of 
Brownian dynamics simulations. Indeed, as $\lambda_H {\dot \gamma}$ 
increases for fixed values of $z$ and $\mu$, both the Gaussian 
approximation, and the first order perturbation expansion become 
increasingly accurate. In particular, if both the approximations 
are accurate at zero shear rate, they continue to remain accurate at
non-zero shear rates. This can be understood in the case of the 
first order perturbation expansion by considering 
equations~(\ref{etaper}) and~(\ref{Psi1per}). Clearly, the departure from 
the $\theta$-solvent values decreases as $\lambda_H {\dot \gamma}$ 
increases. This is inline with the intuitive expectation of 
decreasing excluded volume interactions with increasing shear rate. 

For a fixed value of the shear rate $\lambda_H {\dot \gamma}$, 
as $z$ increases, or $\mu$ decreases, the predictions of the 
Gaussian approximation and the first order perturbation expansion 
become increasingly inaccurate,
with the first order perturbation expansion breaking down before 
the Gaussian approximation. This can be expected, since the 
Gaussian approximation---being exact to first order in $z$---is 
at least as accurate as the first order perturbation expansion. 

The two different Brownian dynamics simulation algorithms mentioned 
in section~5.2 were used to obtain the data in figures~\ref{fig5} 
and~\ref{fig6}. While the algorithm with variance reduction 
was used for shear rates upto $\lambda_H {\dot \gamma} = 0.1$, the 
algorithm without variance reduction 
was used at higher shear rates. With regard to the results obtained by 
variance reduction, it was found that the variance was typically reduced
by a factor of five to ten by the parallel equilibrium simulation
subtraction procedure. While the magnitude of the reduced variance 
was relatively independent of shear rate for the viscosity, it decreased
with increasing shear rate for the first normal stress difference. 
The time to reach steady-state, from start-up of flow, was roughly ten 
relaxation times for $z=0.1$, $z=3$ and $z=30$, and roughly fifteen  
relaxation times for $z=100$. Rheological properties in the two parallel 
simulations remained correlated during the time required to reach
steady-state, and as a result the present technique proved adequate
for the purpose of variance reduction. 

\begin{figure}[!htb] \centerline{ \epsfxsize=4in \epsfbox{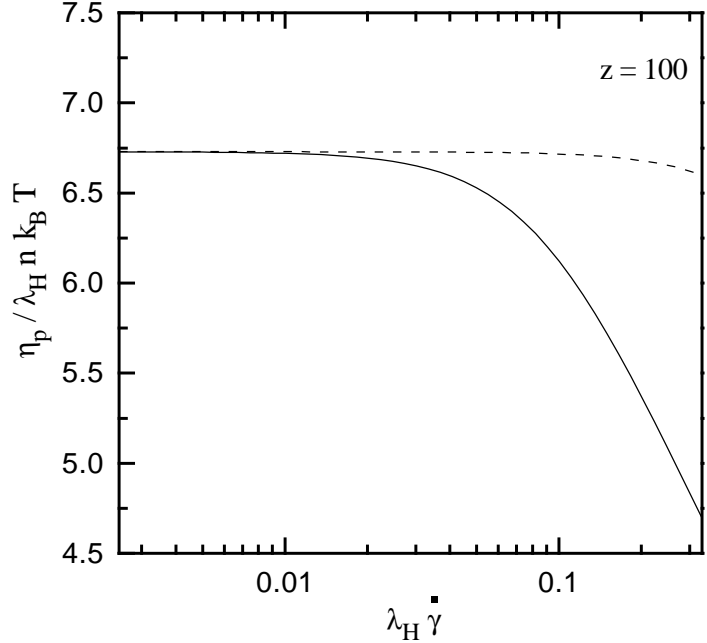}}
\caption{ \footnotesize Non-dimensional viscosity versus
non-dimensional shear rate for $z=100$.  The continuous line is the
prediction of Fixman's theory, and the dashed line is the prediction of
the Gaussian approximation for $\mu=0$. } \label{fig7} \end{figure}

We have seen earlier---in figure~\ref{fig1}---that both Fixman's theory, and
the Gaussian approximation for $\mu=0$, lead to identical values for the
zero shear rate viscosity. This coincidence is, however, restricted to
the limit of zero shear rate. At non-zero shear rates, as can be seen
from figure~\ref{fig7}, there is considerable divergence between the
predictions of the two theories.

\begin{figure}[!htb] \centerline{ \epsfxsize=4in \epsfbox{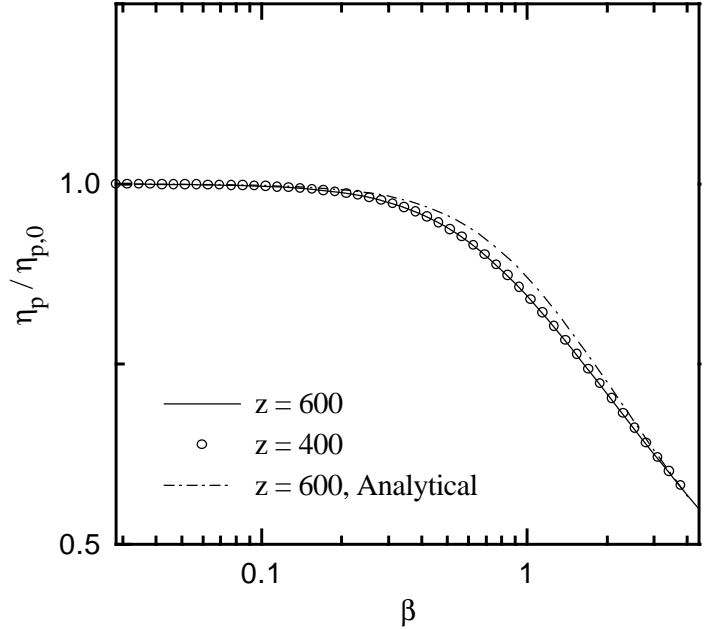}}
\caption{ \footnotesize Reduced viscosity versus reduced shear rate
predicted by Fixman's theory for large values of $z$.  The continuous
line and circles are obtained numerically, while the dot-dashed line is
obtained with the analytical expression~(\ref{anal}). } \label{fig8}
\end{figure}

Fixman's theory for dumbbells, though differing considerably 
from the narrow Gaussian potential in terms of its predictions of 
rheological properties, has the appealing aspect that it captures some of
the {\it universal} features of the behavior of good solvents---which
can only be expected from bead-spring chain theories in the limit of a
large number of beads. We have seen this universal behavior earlier  
in the correct prediction of the end-to-end distance scaling, and in 
the parameter free nature of the ratio $U_{\Psi \eta}$. 
Figure~\ref{fig8} displays the
prediction by Fixman's theory of the reduced variable $(\eta_p /
\eta_{p,0})$ versus the non-dimensional shear rate $\beta = \lambda_p
{\dot \gamma},$ where, $\lambda_p = ( \, \lbrack \eta \rbrack_0 \, M \,
\eta_s / N_{\rm A} \, k_{\rm B}\, T \, ) $, is a characteristic
relaxation time. Here, $\lbrack \eta \rbrack_0 $ is the zero shear rate
intrinsic viscosity, $M$ is the molecular weight and $ N_{\rm A}$ is
Avagadro's number.  For dilute solutions one can show that, $\beta =
\eta_{p,0} \, {\dot \gamma} / n \, k_{\rm B}\, T$. The figure clearly
reveals that as $z \to \infty$, the curves for different values of $z$
overlap. Therefore, in this respect also, Fixman's theory mimics the
universal behavior expected of long chains. The source of this behavior
can be understood by examining equation~(\ref{sfs2}). In the limit of
$\alpha \gg 1$, one can show that $s_2 = [\alpha^6 \, (\lambda_H {\dot
\gamma})^{-2}]^{1/5}$. As a result, $(\eta_p / \eta_{p,0}) = [\alpha^2
\, \lambda_H {\dot \gamma}]^{-2/5}$ ---leading to the observed
scaling.  At small values of $\lambda_H {\dot \gamma}$, $(\eta_p /
\eta_{p,0}) \to 1$.  One can therefore construct the analytical
expression, \begin{equation} {\eta_p \over  \eta_{p,0}} = (1 + \alpha^4
\lambda_H^2 {\dot \gamma}^2)^{-1/5} \label{anal} \end{equation} and
expect it to be accurate at very small and large values of $\beta$. The
dot-dashed curve in figure~\ref{fig8} shows that equation~(\ref{anal})
is accurate over a fairly wide range of $\beta$.

The universal behavior discussed above is not exhibited by the Gaussian
approximation.  The reason for this can be easily understood in the
limit $\mu \to 0$, where, the second moment equation~(\ref{gasecmom})
reduces at steady-state to a non-linear algebraic equation. Indeed, one
can show that for $\alpha \gg 1$, $(\eta_p / \eta_{p,0}) = [1+
\lambda_H^2 {\dot \gamma}^2]^{-1/5}$. As a result, the normalised
material functions do not collapse onto a single curve when plotted
versus $\beta$. It is, however, not realistic to expect a dumbbell
model to exhibit universal features, and the real verification of
universal behavior requires the development of a theory for long
bead-spring chains.

The non-dimensional ratio $U_{\Psi \eta}$ [see equation~(\ref{upsieta})] 
has a constant value of two, independent of shear rate, for 
Hookean dumbbells in $\theta$-solvents, and in Fixman's theory 
for good solvents. Figure~9 displays the predictions of 
$U_{\Psi \eta}$ by the narrow Gaussian potential, obtained by 
Brownian dynamics simulations, the Gaussian approximation, and 
the following first order perturbation expansion (which can be 
derived from equations~(\ref{etaper}) and~(\ref{Psi1per})), 
\begin{equation}
U_{\Psi \eta} = 2  - 2 \, z \,  {1 + 
\lambda_H^2 {\dot \gamma}^2 \over 
\sqrt{1 + \mu^2 } \, \Delta^{3/2} } 
\label{upsietaper}  
\end{equation}  
where, $\Delta$ has been defined below equation~(\ref{Psi1per}).
Since a logarithmic scale has been chosen for the shear rate axis,
it is difficult to represent the zero shear rate value of 
$U_{\Psi \eta}$. However, since it is very nearly constant at low 
values of shear rate, the zero shear rate value is represented in 
figure~9 by the filled circles on the $y$-axis. All the data in 
figure~9 have the same trend of remaining nearly constant at 
low shear rates, and approaching asymptotically the value of two at 
high shear rates. In the case of the first order perturbation 
expansion, this can be understood by considering 
equation~(\ref{upsietaper}) in the limit $\lambda_H {\dot \gamma} 
\to \infty$. 
  
\begin{figure}[!htb] 
\centerline{ \epsfxsize=4in \epsfbox{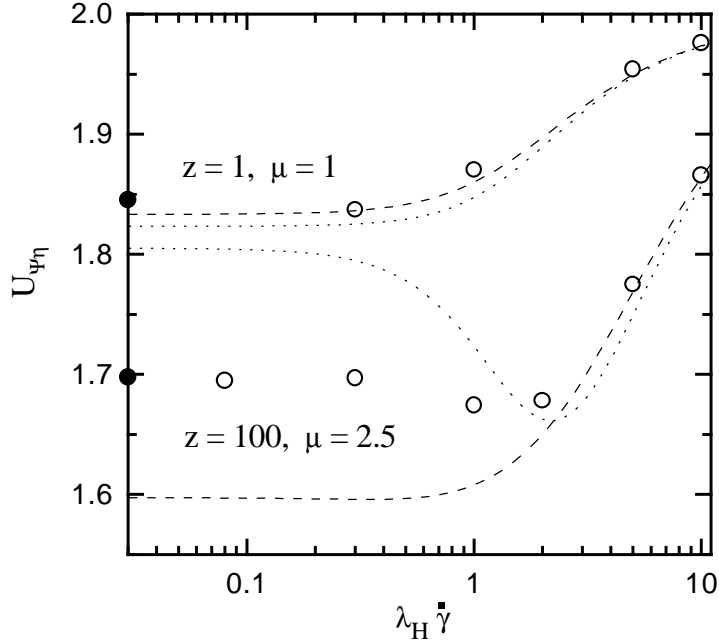}}
\caption{ \footnotesize The ratio $U_{\Psi \eta}$ [see 
equation~(\ref{upsieta})] versus  $\lambda_H {\dot \gamma}$. 
The circles are results of Brownian dynamics simulations, 
the dashed lines are the predictions of the Gaussian approximation, 
and the dotted lines are the predictions of the first order 
perturbation expansion. The filled circles on the $y$-axis represent 
zero shear rate values of $U_{\Psi \eta}$ obtained by equilibrium 
simulations. The error bars in the Brownian dynamics simulations 
are smaller than the size of the symbols.  } 
\label{fig9}
\end{figure}

\begin{figure}[!htb]  
\centerline{\epsfxsize=4in \epsfbox{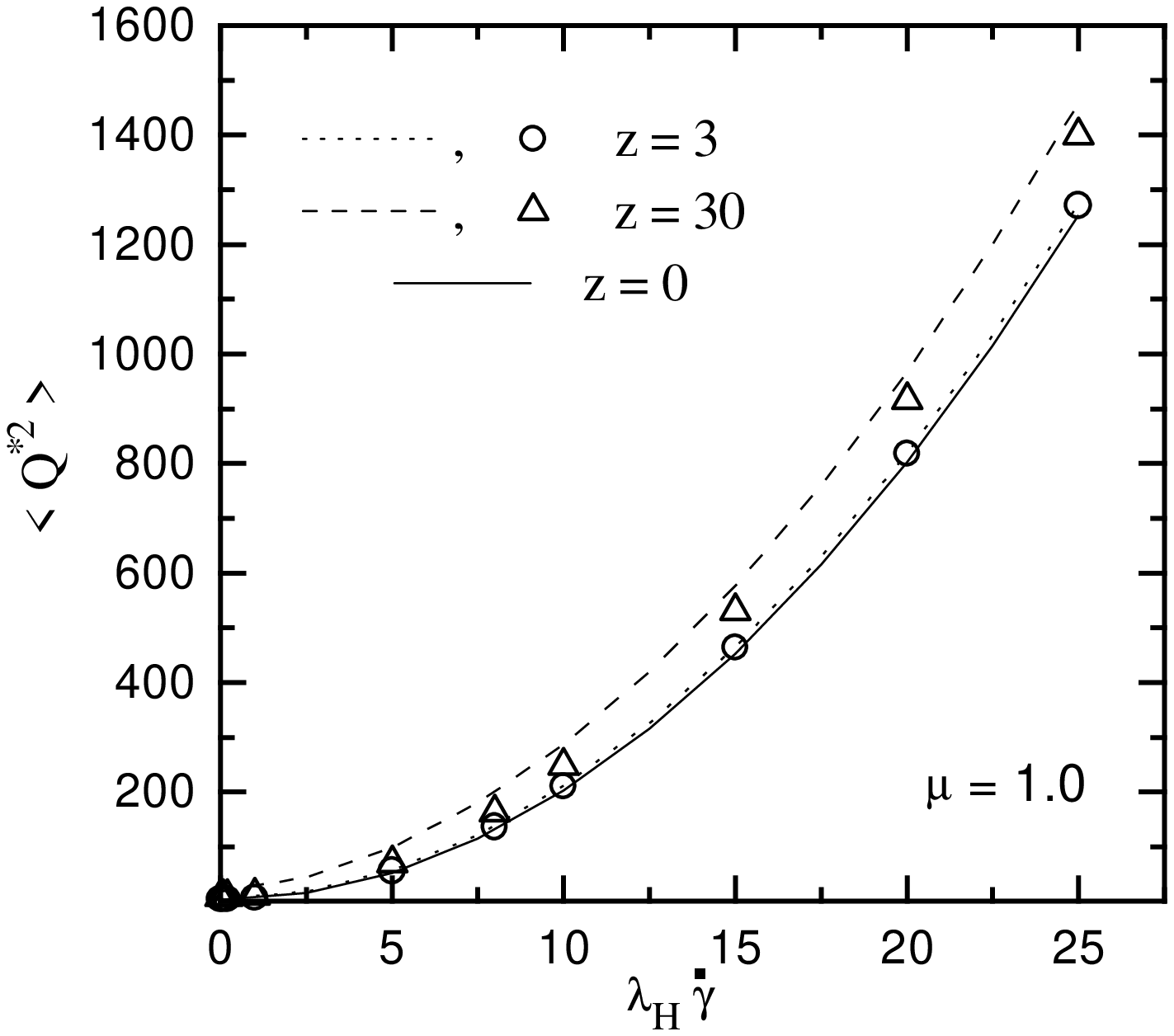}} 
\caption{ \footnotesize 
Non-dimensional mean squared end-to-end vector versus  
non-dimensional shear rate $\lambda_H {\dot \gamma}$,  
for two values of $z$ at $\mu = 1$. 
The circles and triangles are results of Brownian dynamics simulations,  
the dotted and dashed lines are the predictions of the first order  
perturbation expansion, and the continuous line is the analytical 
solution for a theta solvent. The error bars in  
the Brownian dynamics simulations are smaller than the size of the  
symbols. } 
\label{fig10} 
\end{figure} 

The dependence of the mean squared end-to-end distance 
$\avel {Q^*}^{2} \aver$ on 
$\lambda_H {\dot \gamma}$ is revealed in Figure~10.  
The circles and triangles are the results of Brownian dynamics simulations 
obtained with the algorithm without variance reduction. The dotted and 
dashed lines are plots of the following expression, 
\begin{equation} 
\avel {Q^*}^{2} \aver =  3 + 2 \, \lambda_H^2 {\dot \gamma}^2 
+  {z \over \sqrt{1 + \mu^2 } \, \Delta^{3/2} }  
\left[ \, 3(1+\mu^2) + (5+6\mu^2) \, \lambda_H^2 {\dot \gamma}^2 
+2 \, \lambda_H^4 {\dot \gamma}^4   \right] 
\end{equation} 
obtained from the first order perturbation expansion, and  
the continuous line is the well known result for a theta 
solvent~\cite{bird2}, 
$\avel {Q^*}^{2} \aver =  3 + 2 \, \lambda_H^2 {\dot \gamma}^2$. 
Interestingly, the effect of excluded volume interactions on swelling 
increases with increasing shear rate. In the case of the 
first order perturbation expansion---which appears to be accurate for 
$z=3$, but not for $z=30$---this can be understood by 
noting, in the expression above, that the term representing the  
correction to the theta solvent result due to the presence of excluded 
volume increases as the shear rate increases.

\section{Conclusions}

The use of a narrow Gaussian potential, to describe the 
excluded volume interactions between the beads of a Hookean 
dumbbell model, leads to the prediction of swelling and 
shear thinning for relatively 
small non-zero values of the extent of interaction $\mu$. This 
is essentially caused by an increase in the magnitude of 
the equilibrium moments relative to their $\theta$-solvent values. 
A delta function description of the excluded volume potential, 
on the other hand, is found to predict neither swelling nor 
shear thinning for Hookean dumbbells.

For a given strength of the excluded volume interaction $z$, the
Gaussian approximation is found to be reasonably accurate for values of
$\mu$ larger than some threshold value. The behavior of the Gaussian 
approximation can be understood by comparing its predictions with 
those of a first order perturbation expansion in $z$, since it 
is shown here to be exact to first order in  $z$. The perturbation 
expansion reveals that the departure of the viscometric functions 
from their $\theta$-solvent values increases with increasing $z$, 
but decreases with increasing $\mu$, and increasing shear rate 
$\lambda_H {\dot \gamma}$. 

The use of a quadratic potential in Fixman's theory leads to the 
prediction of viscometric functions which are considerably
different from those of the narrow Gaussian potential.
However, Fixman's theory for dumbbells reproduces a number of universal
features observed in good solvents.

\vskip15pt
\noindent
{\bf \large Acknowledgement} 
\vskip15pt
Support for this work through a 
grant III. 5(5)/98-ET from the Department of Science and Technology, 
India, to J. Ravi Prakash is acknowledged. Part of this work was carried out 
while JRP was a participant in the research programme 
{\em Jamming and Rheology} at the Institute for Theoretical Physics, 
University of California, Santa Barbara, USA. JRP would also like to 
thank the non-linear dynamics group at IIT Madras for providing 
the use of their computational facility.

\appendix

\section{The uniform expansion model}

The equilibrium average of any quantity $X(\bQ)$ is given by,
\begin{equation} \avel X \aver_{\rm eq}= {\cal N_{\rm eq}} \, \int X
(\bQ) \, \exp \,  \bigl\lbrace \, - \, {3 \over 2 b^2 } \, Q^2  - E^*
\, \bigr\rbrace \, d \bQ \label{xavg} \end{equation} where, $b^2 = 3
(k_B T / H) $.  By multiplying and dividing the integrand with the
Gaussian distribution $\psi_{\rm eq}^\prime (\bQ)$ [defined by
equation~(\ref{uemgau})], and using the normalisation conditions on $
\psi_{\rm eq} (\bQ)$ and $ \psi_{\rm eq}^\prime (\bQ)$,
equation~(\ref{xavg}) can be rewritten in the form, \begin{equation}
\avel X \aver_{\rm eq}=  {  \avel X (\bQ) \, e^{ - B (\bQ) } \aver_{\rm
eq}^\prime \over \avel e^{ - B (\bQ) } \aver_{\rm eq}^\prime }
\label{xavgmod} \end{equation} where, $B(\bQ)= (3 / 2) \, [ \, (1 /
b^2) - (1 /{b^\prime}^2 ) \, ] \, Q^2 +  E^*$.  For $ \psi_{\rm
eq}^\prime (\bQ) $ to be a good approximation to $ \psi_{\rm eq} (\bQ)
$, $B(\bQ)$ must be small. On expanding $\exp ( - B (\bQ) ) $ in a
Taylor's series, one can then write equation~(\ref{xavgmod}), to first
order in $B(\bQ)$ as, \begin{equation} \avel X \aver_{\rm eq}=\avel X
\aver_{\rm eq}^\prime - \avel X \, B \aver_{\rm eq}^\prime + \avel X
\aver_{\rm eq}^\prime \, \avel B \aver_{\rm eq}^\prime \end{equation}
If $b^\prime$ is chosen such that $\avel X \, B \aver_{\rm eq}^\prime =
\avel X \aver_{\rm eq}^\prime \, \avel B \aver_{\rm eq}^\prime,$ it
then follows that $\avel X \aver_{\rm eq}$ is well approximated by
$\avel X \aver_{\rm eq}^\prime$.

The choice of $b^\prime$ that best optimises the approximation clearly
depends on the quantity $X(\bQ)$ that is averaged.  In our case we
require averages of the quantities $X_{(m)} \equiv Q^{2  m}, \, m=
1,2,3$. From the well known result for the moments of a Gaussian
distribution, \begin{equation} \avel Q^{2  m} \aver_{\rm eq}^\prime =
{2 \over \sqrt{\pi} } \, \biggl[ \, {2 {b_{(m)}^\prime}^2 \over 3} \,
\biggr]^m \, \Gamma (m + {3 \over 2}) \end{equation} where,
${b_{(m)}^\prime}$  represents the value of $b^{\prime}$ corresponding
to $ Q^{2  m}$.  It follows that the optimum value of $b_{(m)}^\prime$
is found by solving the following non-linear algebraic equation for
$\alpha_{(m)}$, \begin{equation} (\alpha_{(m)}^2 - 1) \, m + { z \over
[ \alpha_{(m)}^2 + \mu^2 ]^{3/2} } \, \bigl\lbrace \, { \mu^{2m} \over
[ \alpha_{(m)}^2 + \mu^2 ]^m } - 1 \, \bigr\rbrace = 0 \end{equation}
where, $\alpha_{(m)} =  {{b_{(m)}^\prime} / b} \, ; \, m= 1,2,3$.  In
terms of $\alpha_{(m)}$, \begin{equation} u_m =  \alpha_{(m)}^{2m} \,
\prod_{p=1}^m (2p+1) \end{equation}

\end{document}